\documentclass[11pt,a4paper,total={6in, 8in}]{article}
\usepackage{graphicx,amssymb,amsmath,color}
\usepackage[linkcolor={blue},citecolor={red},colorlinks=true]{hyperref}
\usepackage{cite}
\usepackage{relsize}
\usepackage{enumerate}
\usepackage[margin=2cm]{geometry}
\usepackage{slashed}
\usepackage{multirow}
\usepackage[normalem]{ulem}
\usepackage[dvipsnames, x11names]{xcolor}
\usepackage[compat=1.0.0]{tikz-feynman}
\usepackage{xcolor}
\usepackage{mathtools}
\usepackage{cancel}
\usepackage{soul}
\usepackage{amsmath}

\def\g{\gamma}

\def\beq{\begin{equation}}
\def\eeq{\end{equation}}
\def\bea{\begin{eqnarray}}
\def\eea{\end{eqnarray}}

\def\bit{\begin{itemize}}
\def\eit{\end{itemize}}

\def\l{\left}
\def\r{\right}

\def\baa{\begin{array}}
\def\eaa{\end{array}}

\def\simgt{\mathrel{\lower2.5pt\vbox{\lineskip=0pt\baselineskip=0pt
           \hbox{$>$}\hbox{$\sim$}}}}
\def\simlt{\mathrel{\lower2.5pt\vbox{\lineskip=0pt\baselineskip=0pt
           \hbox{$<$}\hbox{$\sim$}}}}
\newcommand{\vev}[1]{ \langle {#1} \rangle }

\def\bfc{\begin{figure}\begin{center}}
\def\efc{\end{center}\end{figure}}
\def\nn{\nonumber\\}

\newcommand{\blue}[1]{{\color{blue} #1}}

\definecolor{chromeyellow}{rgb}{1.0, 0.65, 0.0}
\definecolor{darkcoral}{rgb}{0.8, 0.36, 0.27}
\definecolor{cadmiumgreen}{rgb}{0.0, 0.42, 0.24}

\begin{document}

\begin{flushright}
\hspace{3cm} 
SISSA  11/2024/FISI
\end{flushright}

\vspace{.6cm}
\begin{center}

\hspace{-0.4cm}{\Large \bf 
Populating secluded dark sector with ultra-relativistic bubbles \\}

\vspace{1cm}{Aleksandr Azatov$^{a,1}$, Xander Nagels$^{b,2}$, Miguel Vanvlasselaer$^{c,3}$, Wen Yin$^{d,4,5}$ }
\\[7mm]
\end{center}

\vskip 0.4cm
\begin{center}
{\hspace{-.08cm}\it $^1$ SISSA International School for Advanced Studies, Via Bonomea 265, 34136, Trieste, Italy}\par
\vskip0.2cm
{\it $^2$ INFN - Sezione di Trieste, Via Bonomea 265, 34136, Trieste, Italy}\par
\vskip0.2cm
{} \par
\vskip0.2cm
{\it $^3$ Theoretische Natuurkunde and IIHE/ELEM, Vrije Universiteit Brussel,
\& The International Solvay Institutes, Pleinlaan 2, B-1050 Brussels, Belgium }\vskip0.2cm
{\it $^4$ Department of Physics,
\&   Tokyo Metropolitan University, Tokyo 192-0397, Japan}
\vskip0.2cm{\it $^5$ Department of Physics,
\&   Tohoku University, Sendai  Miyagi 980-8578, Japan}

\vskip1.cm
\end{center}

\bigskip \bigskip \bigskip

\centerline{\bf Abstract} 
\begin{quote}

 We study Dark Matter  production during 
first order phase transitions from bubble-plasma 
collisions. We focus on scenarios where the Dark Matter sector is secluded and its interaction with the visible sector (including the Standard Model) originates 
from dimension-five and dimension-six operators. We find that such DM is generally heavy and has a large initial velocity, leading to the possibility of  DM
being warm today.
We differentiate between the cases 
of weakly and strongly coupled dark sectors,
where, in the latter case, we focus on glueball DM, which turns out to have very distinct phenomenological properties. We also systematically compute the Freeze-In production of the dark sector and compare it with the bubble-plasma DM abundances.

\end{quote}

\vfill
\noindent\line(1,0){188}
{\scriptsize{ \\ E-mail:
\texttt{$^a$\href{mailto:aleksandr.azatov@NOSPAMsissa.it}{aleksandr.azatov@sissa.it}},
\texttt{$^b$\href{Xander.Staf.A.Nagels@vub.be}{Xander.Staf.A.Nagels@vub.be},
\texttt{$^c$\href{miguel.vanvlasselaer@NOSPAMsissa.it}{miguel.vanvlasselaer@sissa.it}}}
\texttt{$^d$\href{wen@tmu.ac.jp}{wen@tmu.ac.jp}}}
}

\newpage

\newpage

\section{Introduction}

Phase transitions (PTs) taking place in the early universe, often referred to as \textit{cosmological} phase transitions, received a continuous and growing attention. From a theoretical perspective, since the zero temperature potential is expected to be a complicated manifold with several minima, PTs are considered to be common phenomena within quantum field theory. From a phenomenological perspective, PTs might exhibit intriguing implications for the thermal history of the early universe. In particular, a first order PT (FOPT), where a transition occurs from a metastable to a more stable vacuum state, i.e. to a deeper minimum of the potential, might lead to many phenomenological consequences such as baryogenesis \cite{Kuzmin:1985mm, Shaposhnikov:1986jp,Nelson:1991ab,Carena:1996wj,Cline:2017jvp,Long:2017rdo,Bruggisser:2018mrt,Bruggisser:2018mus,Morrissey:2012db,Azatov:2021irb, Huang:2022vkf, ,Baldes:2021vyz, Chun:2023ezg}, the production of heavy dark matter\cite{Falkowski:2012fb, Baldes:2020kam,Hong:2020est, Azatov:2021ifm,Baldes:2021aph, Asadi:2021pwo, Lu:2022paj,Baldes:2022oev, Azatov:2022tii, Baldes:2023fsp,Kierkla:2022odc, Giudice:2024tcp}, primordial black holes\cite{10.1143/PTP.68.1979,Kawana:2021tde,Jung:2021mku,Gouttenoire:2023naa,Lewicki:2023ioy} and possibly observable gravitational waves (GWs)\cite{Witten:1984rs,Hogan_GW_1986,Kosowsky:1992vn,Kosowsky:1992rz,Kamionkowski:1993fg, Espinosa:2010hh}. In a related way, FOPTs occur naturally in a large variety of motivated BSM models like composite Higgs\cite{Pasechnik:2023hwv, Azatov:2020nbe,Frandsen:2023vhu, Reichert:2022naa,Fujikura:2023fbi}, extended Higgs sectors\cite{Delaunay:2007wb, Kurup:2017dzf, VonHarling:2017yew, Azatov:2019png, Ghosh:2020ipy,Aoki:2021oez,Badziak:2022ltm, Blasi:2022woz,Banerjee:2024qiu}, axion models\cite{DelleRose:2019pgi, VonHarling:2019gme}, dark Yang-Mills sectors\cite{Halverson:2020xpg,Morgante:2022zvc} and $B-L$ breaking sectors\cite{Jinno:2016knw, Addazi:2023ftv}.

The interactions between the bubble wall and the plasma have recently attracted much attention. In the regime of relativistic bubble expansion (BE), when the boost factor $\gamma_w  \equiv 1/\sqrt{1-v_w^2}\gg 1$ ($v_w$ is the velocity of the wall), it was first shown in \cite{Bodeker:2017cim} 
that the ultra-fast bubble wall could allow exotic $1 \to 2$ interactions, otherwise forbidden in a Lorentz-invariant background. Subsequently, \cite{Azatov:2020ufh} argued that particles much heavier than the scale of 
the transition could be produced in $1 \to 1$ and $1 \to 2$ processes due to the Lorentz violating bubble wall background and will propagate in shells around  
the bubble wall. A broad review of the different particle production mechanisms and their corresponding interactions with the bubble wall is given in \cite{Baldes:2024wuz}. 

It was later suggested \cite{Azatov:2021ifm} that the special class of $1 \to 2$ processes could leads   to the production of heavy scalar Dark Matter (DM) via the operator $\lambda \phi^2 h^2$, where $h$ is the field undergoing the phase transition, potentially playing the role of the Higgs or another scalar field, and $\phi$ is the heavy DM particle. Due to the large boost factor, reached by ultra-relativistic or runaway bubble walls, DM produced in $h \to \phi \phi$ transitions will be strongly boosted with an average energy in the plasma frame given by $\bar E_{\phi, \rm plasma} \sim M_\phi^2/T_{\rm nuc}$, where $T_{\rm nuc}$ is the nucleation temperature and $M_\phi$ the mass of the scalar $\phi$. Based on this realisation, authors of\cite{Baldes:2022oev} proposed that the bubble wall production mechanism could induce \emph{heavy and warm}  Dark Matter (WDM), using again an interaction of the form $\lambda \phi^2 h^2$.

Nevertheless, the DM sector is not necessarily scalar by nature or required to share a renormalizable interaction with the phase transition sector, containing the $h$ field (which in principle could be related to the SM Higgs or not).  If the DM sector does not share renormalisable interactions with the SM, it is said to be \emph{secluded}. Such secluded sectors typically interact with the thermal bath via non-renormalisable operators with a characteristic scale $\Lambda_{}$, of the type 
\bea 
\label{eq:higher_dim}
\frac{h^2 {\cal O}_{\rm DS}}{\Lambda^{d-2}} \, ,
\eea 
with ${\cal O}_{\rm DS}$ being a function of fields with dimension $d$, containing the DM candidate. In this paper, we study the production of heavy and potentially warm DM, generated by a phase transition via such non-renormalisable operators in Eq.\eqref{eq:higher_dim}. We assume that the produced particles constitute the entire DM abundance. Consequently, we will see that the phase transition scalar field $h$ \emph{cannot} be the physical SM-like Higgs if the production mechanism is required to produce the observed abundance of DM, but has to be another scalar field, which may be a $B-L$ Higgs boson for the neutrino mass~\cite{Minkowski:1977sc,Yanagida:1979as,Ramond:1979py,Gell-Mann:1979vob,Mohapatra:1979ia},  a Peccei Quinn Higgs boson~\cite{Kim:1979if,Shifman:1979if,Dine:1981rt,Zhitnitsky:1980tq}, or some Higgs associated with the flat direction for thermal inflation/supercooling ~\cite{Yamamoto:1985rd,Lyth:1995ka}. These Higgs do not have Standard Model gauge charge and it is easy to have an ultra-relativistic bubble wall expansion during the first order phase transition because the friction can be small. Our discussion applies generically, and we denote them  as \emph{BSM Higgs}. In what follows, we will denote the  SM Higgs with a capital letter $H$ and the BSM Higgs with a lowercase $h$.

\begin{figure}
    \centering
    \includegraphics[width=.19\linewidth]{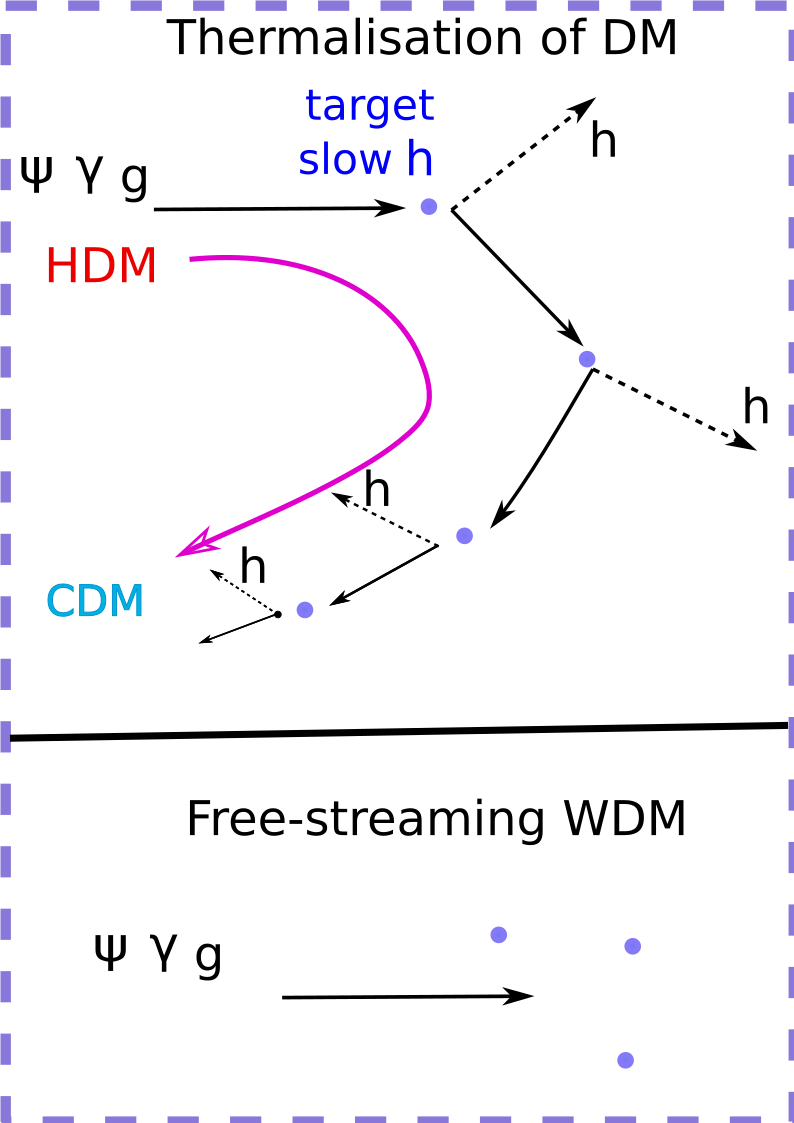}
    \includegraphics[width=.34\linewidth]{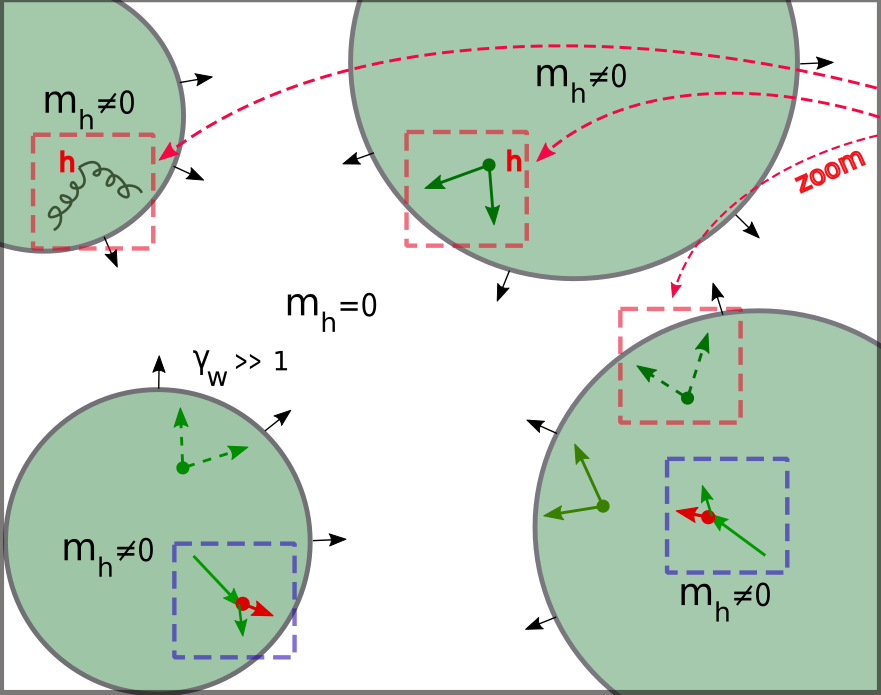}
    \includegraphics[width=.445\linewidth]{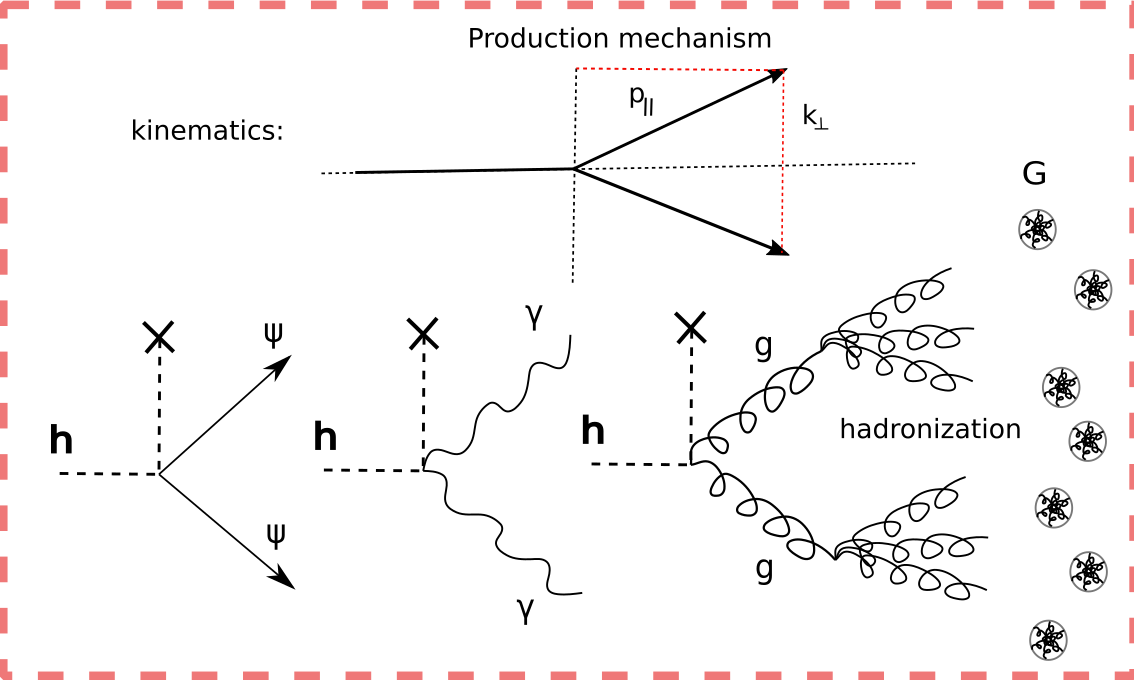}
    
    \caption{Schematic of the production of Dark Matter from bubbles with $k_\perp^2 \sim \gamma_w v T_{\rm nuc}, p_{||} \sim \gamma_w T_{\rm nuc}$. In the middle panel, we show how the expansion of bubble walls can produce very boosted fermions $\psi$, vectors $\gamma$ or gluons $g$ (which become glueballs $G$) depending on the model under consideration. In the right panel, the red rectangle, we present more in details the bubble-plasma production channels. The cross represents the interaction with the bubble wall which allows the DM production. Subsequently, as shown on the left panel, in the blue rectangle, we show the interactions of the boosted produced particles with the thermalised plasma, the blue $h$ particles on the sketch. The fast $\psi, \gamma, g$ can interact with the slow targets $h$ via $h(\psi, \gamma, g) \to h(\psi, \gamma, g)$ and cool down to usual CDM or free-stream and remain warm dark matter. In the main text, we consider either $\psi$, $\gamma$ or $g$ to be produced. }
    \label{Fig:scheme}
\end{figure}

    In this paper, we will consider different natures for the DM particle: DM can be a fermion $\psi$, a dark vector $\gamma$ or  a dark glueball $G$ and we will scrutinize each of these cases  in detail.
Furthermore, we compute the  spectrum of the DM right after bubble production, and track its evolution, taking into account 
scatterings with the bath. We present schematically our mechanism in Fig. \ref{Fig:scheme}, where the production is 
sketched inside the red rectangle and the rescattering after production inside the blue one.

Here is the summary of the new results presented in this paper: 
\begin{itemize}
\item We study the production and the abundance of particles originating 
from a higher-dimensional operator of
    dimension five and six. We find the 
    following scalings for the corresponding  number
    densities for the production of the particles in the weakly coupled cases: $n_{\rm dim~ 5} \propto T_{\rm nuc}^3v^2/\Lambda^2$, $n_{\rm dim ~6} \propto \gamma_w T_{\rm nuc}^4v^3/\Lambda^4$. For the strongly interacting case (glueball DM production), it is more sensible to describe the process in terms of the total energy transferred to the dark sector and we find that the glueball 
    energy density scales as $\rho_G \propto \gamma_w^2 T_{\rm nuc}^4v^4/\Lambda^4$. The scaling of the energy density of the gluons, with the large boost factor, comes from the fact that both the number density of gluons 
    $n_g \propto \gamma_w$ and the energy in the plasma frame of the emitted gluon, $\bar E_g \propto \gamma_w$ scale like $\gamma_w$. 
   
\item For weakly interacting boosted particles, we compute the average velocity at matter-radiation equality and the spectrum of the DM after production, as well as its evolution until matter-radiation equality. 
Bubble wall production appears to be an efficient mechanism to produce heavy warm DM in both the fermion and the vector production.
\item We compute the pressure due to this particle production, which is not only inevitable in our mechanism, but might also be a rather generic effect for FOPTs with fast bubble walls. We find the following relations:    
$\mathcal{P}_5 \propto \gamma_w v^3 T_{\rm nuc}^3/\Lambda^2$ and $\mathcal{P}_6 \propto \gamma_w^2 v^4 
T_{\rm nuc}^4/\Lambda^4$. We however observe that this pressure is parametrically smaller than the leading order $v^2T_{\rm}^2$ pressure over the whole range where our computation holds. 
\item We compute the Freeze-In (FI) abundance produced via the three operators mentioned above and compare with the bubble production via the same operators.
\end{itemize}

The remainder of this paper is organised as follows: in section \ref{sec:reminder}, we remind the reader of the main results obtained in \cite{Azatov:2021ifm, Baldes:2022oev} and highlight the salient characteristics of the scalar emission. In section \ref{sec:weakly_prod}, we present the study of the production of a weakly coupled secluded sector via dimension five and six operators. In section \ref{sec:prod_strongly}, we study the production of high-energy gluons and their confinement leading to glueballs.

\section{Reminder of the production via the renormalisable operator }
\label{sec:reminder}

 To make our story complete, we first review the computation of the production via the \emph{renormalisable} interaction $\lambda \phi^2 h^2$, as it was initially proposed in \cite{Azatov:2021ifm} and then further studied in \cite{Baldes:2022oev}. 
 In this section we consider the following Lagrangian 
 \bea 
 \mathcal{L} = \frac{1}{2}(\partial_\mu \phi)^2 - \frac{1}{2} M_\phi^2 \phi^2 - \frac{\lambda}{2}h^2 \phi^2  \, .
 \eea 
During the phase transition, the BSM Higgs field $h \to h+v$, induces a three-leg vertex in the Lagrangian 
\bea 
\mathcal{L} \subset \lambda v h \phi \phi \quad,
\eea
allowing for splittings $h \to \phi \phi$. Note that this transition would be forbidden in vacuum and only occurs thanks to the bubble wall presence, which breaks Lorentz invariance, leading to the non-conservation of $z$-momentum. Using the WKB approximation, the transition from light to heavy states $h \to \phi\phi$ has a probability of the form\cite{Azatov:2021ifm}\footnote{Notice the factor of two difference with \cite{Azatov:2021ifm}.}
\bea
\label{eq:prod_scalar}
P_{h \to \phi^2} \approx \bigg(\frac{\lambda v}{M_\phi}\bigg)^2\frac{1}{48\pi^2} \Theta ( 1- \Delta p_z L_w) \simeq \bigg(\frac{\lambda v}{M_\phi}\bigg)^2\frac{1}{48\pi^2} \Theta \l( p_0-\frac{2M_\phi^2}{v} \r).
\eea
Where $L_w$ is the width of the wall, which is approximately  $L_w
 \sim 1/v$, with $v \ll M_\phi$, and $\Delta p_z \equiv p^h_z - p^\phi_{z,b}-p^\phi_{z,c} \approx M_\phi^2/(2p^h_z x(1-x)) = 2M_\phi^2/p^h_z$ is the difference of momenta 
 between final- and initial-state particles in the direction 
 orthogonal to the wall, and we took $x = 1/2$ in the last equality.  The $\Theta(1- \Delta p_z L_w)$-function comes from the requirement that the transition is
  \emph{non-adiabatic}. Putting a step cut-off is a rough approximation and the exact behaviour is in principle dependent on the wall shape (see Appendix A of \cite{Azatov:2021ifm} and Appendix H of \cite{Azatov:2024auq} for further discussions). 
  This condition is physically similar to the requirement that the energy in the center-of-mass frame of the collision between a standing particle of the wall and an incoming $h$, which is $s_{\rm prod} \sim 2 p_0 v$, needs to be larger than $4 M_\phi^2$ for on-shell $\phi$ production. \\
In the aftermath of the bubble expansion, a non-thermal 
 abundance of $\phi$ accumulates, which takes the following form 
\bea
n_\phi^{\text{BE, PF}} &\approx & \frac{2}{\gamma_w v_w}  \int \frac{d^3p}{(2\pi)^3} P_{h \to \phi^2} (p) \times f_h (p, T_{\text{nuc}}) 
\nonumber \\ 
&\approx &
 \frac{2\lambda^2 v^2}{48\pi^2 M_{\phi}^2 \gamma_w v_w}  \int \frac{d^3p}{(2\pi)^3}  \times f_h (p,T_\text{nuc})\Theta ( p_z- 2M_\phi^2/v),
\label{eq:density_1}
\eea 
where the subscript PF means evaluated in the plasma frame. 
 $v_w = \sqrt{1-1/\gamma_w^2}$ is the velocity of the wall, and $f_h(p)$ is the equilibrium thermal distribution of $h$ outside of the bubble.  
We assume $h$ to be in thermal equilibrium with the bath at temperature $T_{\rm nuc}$ and is therefore described by a Boltzmann distribution  $f_h (p) \approx e^{-{\gamma_w}(E_h - v_wp^h_z)/T_\text{nuc}}$ with $E_h = \sqrt{p_z^2 + \vec{p}^2_\perp}$. We can thus perform the integral in Eq. \eqref{eq:density_1}, obtaining
\bea
n_\phi^{\text{BE}} =   \frac{\lambda^2 }{96\pi^4 \gamma_w^3 v_w}\times  \frac{v^2T_\text{nuc}^2}{M_\phi^2}\bigg( \frac{M_\phi^2/v}{1-v_w}+ \frac{T_\text{nuc}(2-v_w)}{\gamma_w (v_w-1)^2}\bigg) \times e^{- \gamma_w \frac{2M_\phi^2}{v} \frac{1-v_w}{T_\text{nuc}}}.
\label{eq:density_2}
\eea
With $\gamma_w (1-v_w) = \gamma_w - \sqrt{\gamma_w^2 - 1} \to \frac{1}{2\gamma_w}$  in the limit of fast walls, the density in the plasma frame becomes
\bea
n_\phi^{\text{BE}} &=&  \frac{T_\text{nuc}^3}{24\pi^2} \frac{\lambda^2 v^2}{\pi^2 M_\phi^2}  e^{-  \frac{M_\phi^2}{vT_\text{nuc} \gamma_w }}  + \mathcal{O}(1/\gamma_w) \quad .
\label{eq:density_f}
\eea
 The factor $e^{-  M_\phi^2/(vT_\text{nuc} \gamma_w )}$ is a consequence of $\Theta ( p_0- 2M_\phi^2/v)$ in the the Eq. \eqref{eq:density_1}. We can see that in the non-adiabatic limit,
 \bea
 \gamma_w > \frac{M_\phi^2}{vT_{\text{nuc}}}  \quad ,
 \label{eq:range}
 \eea
 the exponential goes to one and the density becomes independent of the velocity of the wall $v_w$, as opposed to particle production via dimension five and six operators, as we will see below. 
The final {number density} of heavy non-thermal DM, in the unsuppressed region, is of the form 
\bea
n_\phi^{\text{BE, PF}} \approx \frac{\lambda^2 v^2}{ M_\phi^2}\frac{T_\text{nuc}^3}{24\pi^4} e^{-  \frac{M_\phi^2}{vT_\text{nuc} \gamma_w }} \, .
\label{eq:density_fin}
\eea
After the completion of the  PT, the plasma is reheated to some \emph{reheating temperature} $T_{\rm reh}$, that we can compute in the following way 
\bea 
T_{\rm reh} \approx (1+\alpha_{\rm nuc})^{1/4} T_{\rm nuc} \approx v  \,, \qquad \alpha_{\rm nuc} \equiv \frac{\Delta V}{\rho_{\rm rad}}  \, ,
\eea 
where $\Delta V$ is the difference of potential in the broken and the symmetric vacuum. 

Dividing by the entropy density after the PT, $s(T_{\rm reh}) \propto T_{\rm reh}^3$ and redshifting to today, 
the final relic abundance today writes 
\bea
\Omega^{\text{today}}_{\phi,\text{BE}}h^2 \approx {2.7}\times 10^5  \times \bigg(\frac{1}{g_{\star S}(T_{\text{reh}})}\bigg) \bigg(\frac{\lambda^2 v}{ M_\phi}\bigg)\bigg(\frac{v}{\text{GeV}}\bigg)\bigg(\frac{T_\text{nuc}}{T_{\text{reh}}}\bigg)^3e^{-  \frac{M_\phi^2}{vT_\text{nuc} \gamma_w }}.
\label{eq:relic_ab}
\eea
Here, $g_{\star S}(T)$ is the relativistic degrees of freedom for entropy and we will also use $g_{\star}(T)$ to indicate the one for the energy.
Notice that emitted particles are produced with very large boost factor in the plasma frame
\bea
\label{eq:energy_scalar}
\bar{E}_{\phi, \text{plasma}} \approx \frac{1}{2}\frac{
\int{dx\left[(p^0_b+p^0_c) \gamma_w -(p^z_b+p^z_c)v_w \gamma_w \right]}
}{\int{dx}} \sim \frac{1}{2} \frac{M_\phi^2}{T_{\rm nuc}} .
\eea
Here, in the last approximation we have used that $ p^\phi_0\sim \gamma_w (1+v_w )T_{\rm nuc},  v_w= \sqrt{1-\gamma^{-2}_w}$. \\

The authors in Ref. \cite{Baldes:2022oev}  have shown that in 
part of the parameter space of the model DM maintains large velocity apart from 
the usual red-shifting due to the universe expansion and can have a significant \emph{free-streaming} (FS) length $L_{\rm FS}$
\bea
L_{\rm FS}=
\int^{\infty}_{z_{\rm eq}} dz \frac{1}{H}\frac{{V_{\rm eq} \frac{1+z}{1+z_{\rm eq}}}}{\sqrt{\Big(V_{\rm eq}\frac{1+z}{1+z_{\rm eq}}\Big)^2+1}},
\eea 
where we defined $ V_{\rm eq}\equiv V(t_{\rm eq})$ as the average velocity of the DM at matter-radiation equality and $z$ is the redshift. 
Observations of small scale structures constrain $L_{\rm FS}$. 
The strongest constraint for the DM free-streaming comes from Lyman-$\alpha$ for the DM free-streaming length $L_{\rm FS}\lesssim 0.059{\rm Mpc}$, which is recast from sterile neutrino mass bound, $5.3$ keV\cite{Bode:2000gq,Viel:2005qj,Irsic:2017ixq}. 
This leads to the following bound on the average velocity at matter-radiation equality, 
\bea 
V_{\rm eq} 
\lesssim 4.2\times 10^{-5}
\qquad \text{(Lyman-$\alpha$ bound)} \, , 
\label{eq:LyAlphaBoundNow}
\eea 
Similarly, we can recast the future prospects from 21 centimeters (WDM mass $> 14\,$keV with the Hydrogen Epoch of Reionization Array \cite{Sitwell:2013fpa,Munoz:2019hjh}), which leads to 
\bea
L_{\rm FS}<0.018\text{ Mpc}  \qquad \Rightarrow \qquad V_{\rm eq}<1.2\times 10^{-5} \qquad \text{(21 centimeters)} \, ,
\label{eq:21cmBoundFuture}
\eea
and subhalo count (WDM mass $> 18$keV with the Vera Rubin Observatory~\cite{LSSTDarkMatterGroup:2019mwo}) 
\bea
L_{\rm FS}<0.016\text{ Mpc} \qquad \Rightarrow \qquad V_{\rm eq}<1.0\times 10^{-5} \qquad \text{(subhalo count)}.
\eea
The DM particles which have such non-negligible velocities are denoted in the literature as warm (WDM), with  typical velocities
\bea 
V^{\rm WDM}_{\rm eq}\sim 10^{-5} \quad.
\eea 
In traditional mechanisms\cite{Colombi:1995ze} for WDM
the  candidate mass is generally small, around keV mass scale. However the authors of Ref.
\cite{Baldes:2022oev} have shown that the DM produced in the bubble expansion can also be warm, though very heavy. Indeed 
starting from equation (\ref{eq:energy_scalar})
 and assuming that DM is not in kinetic equilibrium with the surrounding plasma we obtain:
 \bea 
\label{eq:velocity_eq}
V_{\rm eq} &&\approx \bigg(\frac{g_{\star,s}(T_{\rm eq})}{g_{\star,s}(T_{\rm reh})}\bigg)^{1/3}\frac{T_{\rm eq} p^i_{\rm DM}}{T_{\rm reh} M_{\rm DM}} \sim 0.3 \frac{T_{\rm eq} p^i_{\rm DM}}{T_{\rm reh} M_{\rm DM}} \nn
&&\approx 0.3 \frac{T_{\rm eq} \bar{E}_\phi}{T_{\rm reh} M_{\phi}}  \approx   10^{-10} \frac{ \text{GeV} \times M_{\phi}}{T_{\rm reh} T_{\rm nuc}} , 
\eea
where $T_{\rm eq}$ is the temperature at matter-radiation equality and $T_{\rm reh}$, as we will describe more in depth later, the temperature after the completion of the PT.
Thus the DM produced in the bubble expansion will be warm if 
\bea 
10^5 v T_{\rm nuc} \sim  M_\phi \times \text{  GeV} \, .
\eea 
Very interestingly, this indicates a viable parameter space for explaining the observed abundance of DM being heavy and warm in a range roughly $v \sim \mathcal{O}(100) \text{GeV}, M_\phi \sim 10^{(8-9)} $ GeV and mild supercooling\cite{Baldes:2022oev}. We now pursue this line of investigation with secluded DM and more realistic DM models. At last we would like to comment on the DM production from the bubble-bubble collisions \cite{Falkowski:2012fb,Mansour:2023fwj,Giudice:2024tcp,Shakya:2023kjf}. This process is always present but is generically subdominant to the DM production \emph{if the non-adiabaticity constraint is satisfied}.  
As a consequence, we will neglect it in the rest of this study.

\section{Production of weakly coupled particles}
\label{sec:weakly_prod}
After the reminder of production via renormalisable operators, we now turn to the production via non-renormalisable operators of the type $ h^2 {\cal O}_{\rm DS}/\Lambda^{d-2}$. In this section, we will study the bubble production of dark matter within the framework of weakly coupled theories, where the DM can be a fermion or a dark vector. We also compute the parameter space where it can constitute the total amount of DM, its average velocity and its spectrum at and after the production.

\subsection{Production from dimension 5 operator and fermion DM}

We start with the case of fermionic DM $\psi$, which carries conserved quantum number responsible for its stability,  coupled to the thermalised sector $h$ via the following non-renormalisable operator 
\bea 
\label{eq:fermion_op}
 \mathcal{L} \supset \frac{h^2 \psi \bar \psi}{\Lambda} \; ,
\eea 
where $\Lambda$ is the UV cutoff and $\psi$ is the fermion DM of mass $M_\psi$. A very similar procedure to the one presented in the previous section \ref{sec:reminder} can be followed for the operator in Eq.\eqref{eq:fermion_op}. We show in Appendix \ref{App:prodDM} that the probability $P_{h \to \psi \psi}$ of this splitting is given by 
 \begin{align}
   P_{h \to \psi \psi}(p_0)&= \frac{v^2}{8\pi^2\Lambda^2} G(p_0) \Theta (p_0  - 2M_\psi^2L_w) \Theta (\Lambda^2-2p_0v ) 
   \, ,
\end{align}
where $p_0$ is the energy of the incoming $h$ in the wall frame. Let us comment on various factors in this equation.
The first $\Theta$-function, is an anti-adiabaticity constraint and can be understood in a similar way as the $\Theta$-function in Eq.\eqref{eq:prod_scalar}. 

 When the energy in the center of mass $s_{\rm prod} \approx 2\gamma_w vT_{\rm nuc} > \Lambda^2$, the EFT breaks down and we expect the production mechanism to become dependent on the explicit UV completion of the model. For this reason, we will require now that 
 \bea
 s_{\rm prod} \approx 2p_0 v \approx 2\gamma_w vT_{\rm nuc}< \Lambda^2 \qquad \text{(EFT validity condition)} \, , 
 \eea 
as ensured by the second $\Theta$-function of \eqref{eq:fermion_op}. The region that does not meet this condition will be referred to as the \emph{EFT breakdown} region. 
The $G(p_0)$ is a dimensionless function entering the production process which has the following form 
\bea 
G(p_0) \equiv \frac{4}{3}\sqrt{1-\frac{2M_\psi^2}{p_0v}}\left(\frac{M_\psi^2}{2p_0v}-1\right)+\log\left[\left|1-\frac{p_0 v}{M_\psi^2}-\sqrt{\left(\frac{p_0 v}{M_\psi^2}-2\right)\frac{p_0 v}{M_\psi^2}}\right|\right] \,. 
\eea 
 After the bubble wall expansion, a non-vanishing abundance of $\psi$ particles has been created. By performing the same analysis as in the renormalisable case, we obtain, in the plasma frame
\begin{align}
n_\psi^{\text{BE}, \text{PF}} 
&\approx g_h\frac{v^2T_{\rm nuc}^2(1+v_w)}{16\pi^4\Lambda^2\g_wv_w} \left(2\frac{M_\psi^2}{v}+\g_wT_{\rm nuc}(2+v_w-v^2_w)\right)G(\gamma_w T_{\rm nuc})e^{-\frac{2M_\psi^2\g_w}{T_{\rm nuc} v}(1-v_w)} \, ,
\end{align}
leading, if $g_h=1$, to the DM fraction today:

\begin{align}
  \Omega^{\text{today}}_{\psi,\text{BE}}h^2 &\approx 5.38 \times 10^8  \frac{n^{\rm BE}_\psi}{g_\star^s T_{\rm reh}^3}\frac{M_\psi}{\rm GeV} \nonumber \\
  &\approx 5.38 \times 10^8 \frac{1}{g_\star^s} \frac{v^2T_{\rm nuc}^3(1+v_w)}{16\pi^4T_{\rm reh}^3\Lambda^2\g_wv_w}  
  \left(2\frac{M_\psi^2}{vT_{\rm nuc}}+\g_w(2+v_w-v^2_w)\right) G(\gamma_w T_{\rm nuc})e^{-\frac{2M_\psi^2\g_w}{T_{\rm nuc}v}(1-v_w)} \, ,
\end{align}
which for $v_w\approx 1$ gives finally
\begin{align}
\label{eq:density_today_BE}
\Omega^{\text{today}}_{\psi,\text{BE}}h^2 &\approx 1.38 \times 10^6 \left(\frac{1}{g_\star^s} \right)\left(\frac{v}{\Lambda}\right)^2 \left(\frac{T_{\rm nuc}}{T_{\rm reh}}\right)^3\left(\frac{M_\psi}{\rm GeV} \right)\left(\frac{M_\psi^2}{\g_wvT_{\rm nuc}}+1\right) G(\gamma_w T_{\rm nuc})e^{-\frac{M_\psi^2}{T_{\rm nuc }v\g_w}} \,  .
\end{align}

On the top of this production by bubble expansion, there will be a contribution from FI after reheating, computed in Appendix \ref{App:FI_abundances} if $M_\psi \gg T_{\rm reh}$, 
\bea 
\label{eq:density_today_FI}
\Omega^{\rm today}_{\psi, \rm FI} h^2  
\approx 5.84 \times 10^4 \bigg(\frac{M_\psi}{\text{GeV}} \bigg) \frac{ M_{\rm pl} M_\psi }{ g_\star^{3/2} \Lambda^2  } \left(\frac{ T_{\rm reh}}{M_\psi}\right)^{3/2} e^{-2M_\psi/T_{\rm reh}} \, ,
\eea 
which can be sizable if $M_\psi \lesssim 20T_{\rm reh}$. 
Here  $M_{\rm pl} \approx 1.2 \times 10^{19}$ GeV is the Planck mass.
As a consequence, FI production will be subdominant if 
\bea 
\Omega^{\rm FI}_{\psi} h^2 \ll 0.1  \, ,\qquad  M_\psi/T_{\rm reh} > 25 + \log \bigg(\frac{v}{\Lambda } \frac{M_\psi^2}{ v^2} \bigg) \, ,
\eea 
where we approximated $T_{\rm reh} \sim v$ in the argument of the logarithm.\footnote{We also have a contribution of the thermal production before the phase transition.
This can be estimated by replacing $T_{\rm reh}$ in $\Omega^{\rm FI}_{\psi}$ with the reheating temperature by inflaton decay, $T_{\rm R}$, and multiplying an entropy dilution factor $T_{\rm nuc}^3/T_{\rm reh}^3,$ assuming again that $M_{\psi}$ is larger than $T_{\rm R}$.   
We get
\bea 
\Omega^{\rm today}_{\psi, \rm FI} h^2  
\approx 5.84 \times 10^4 \bigg(\frac{M_\psi}{\text{GeV}} \bigg) \frac{ M_{\rm pl} M_\psi }{ g_\star^{3/2} \Lambda^2  } \left(\frac{ T_{\rm R}}{M_\psi}\right)^{3/2}  \left(\frac{ T_{\rm nuc}}{T_{\rm reh}}\right)^{3}e^{-2M_\psi/T_{\rm R}}
\eea
Requiring this to be much smaller than 0.1, we get 
\bea 
 M_\psi/T_{\rm R} \gtrsim 20.
\eea} 
One finds that in our scenario, the upper bound of the reheating temperature, $T_{\rm R}$, before the phase transition cannot be too larger than the upper bound of $T_{\rm reh}$ by the phase transition. Conventionally, we call $T_{\rm reh}$ the reheating temperature after the phase transition, and $T_R$ the temperature of the universe after the reheating due to inflation.

After the DM production, the future of the emitted particles depends on the interactions with the thermal bath, mostly with the $h$ particles, via $\psi h \to \psi h$. There will be two different regimes that we will now study in detail. 

\paragraph{Free-streaming region}
The first possibility is that the interaction $\psi h \to \psi h$ is always out of equilibrium after the PT and  cannot modify the velocities of DM particles. This is the case when 
\bea 
\label{eq:decoupled}
\Gamma_{\psi h \to \psi h} \sim \frac{T_{\rm reh}^3}{8\pi^3 \Lambda^2} \ll H(T= T_{\rm reh}) \, ,\qquad \frac{T_{\rm reh} M_{\rm pl}}{ \Lambda^2} < 1.66 g_\star^{1/2} (8\pi^3)\, .
\eea 
Immediately after the phase transition the average  energy of the $\psi$ field can be approximated as (see Appendix \ref{app:spectrum_at_em} for the computation)

   \bea
 \label{eq:average_E_kperp}  
\bar E_{\psi,\rm plasma}\simeq \frac{L_w^{-1} \gamma_w}{3\log \frac{\gamma_w T}{M_\psi^2 L_w} -5.92} \,. 
   \eea 
We compare this expression with the energy of the DM for the renormalizable portal
in Eq.\eqref{eq:energy_scalar}, the dimension-four scalar portal. We observe that the dimension-five case predicts more energetic DM, with the ratio of energies scaling as:
\bea
\frac{\Bar{E}_{\rm dim~5}}{\Bar{E}_{\rm dim~4}}\sim \frac{\gamma_w T_{\rm nuc} v}{M_{\rm DM}^2}.
\eea

\begin{figure}
    \centering
    \includegraphics[width=.3\linewidth]{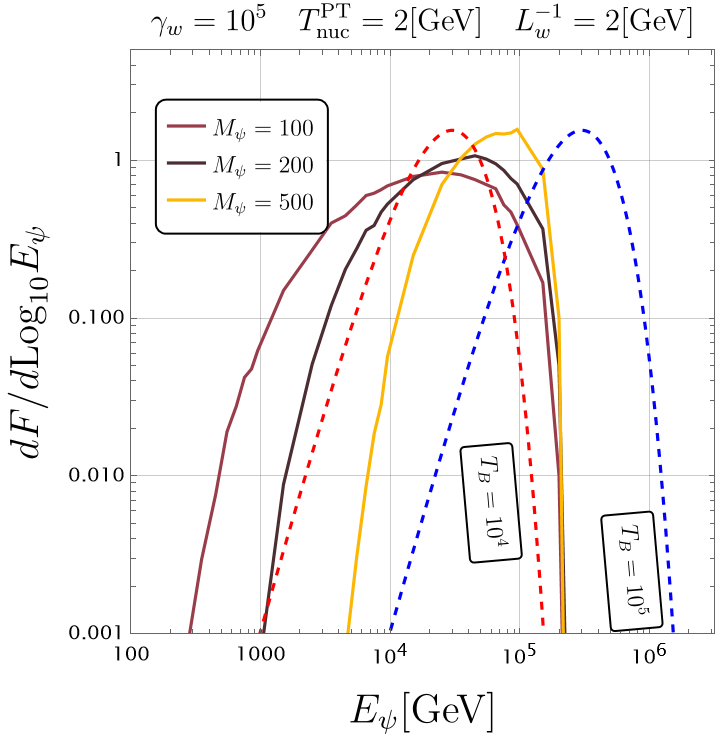}
    \includegraphics[width=.3\linewidth]{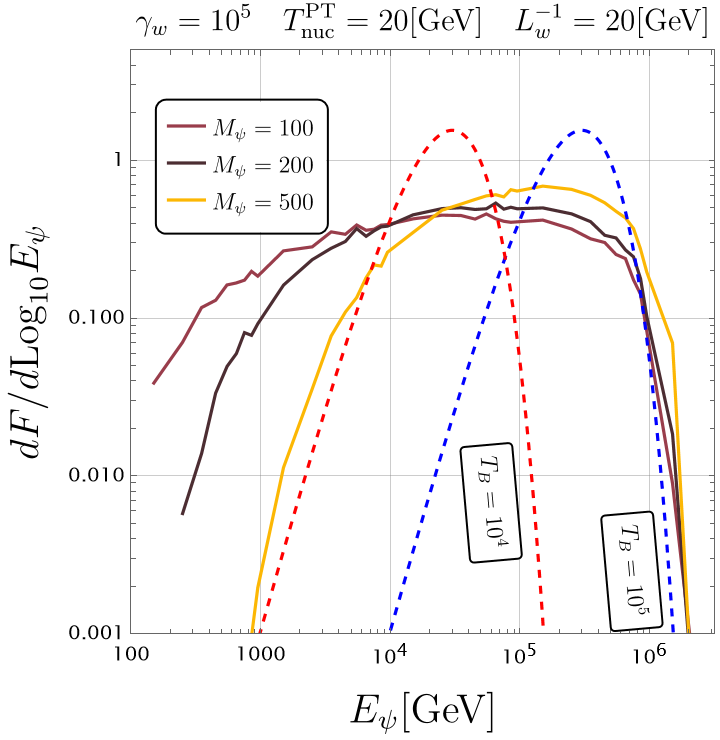}
    \includegraphics[width=.3\linewidth]{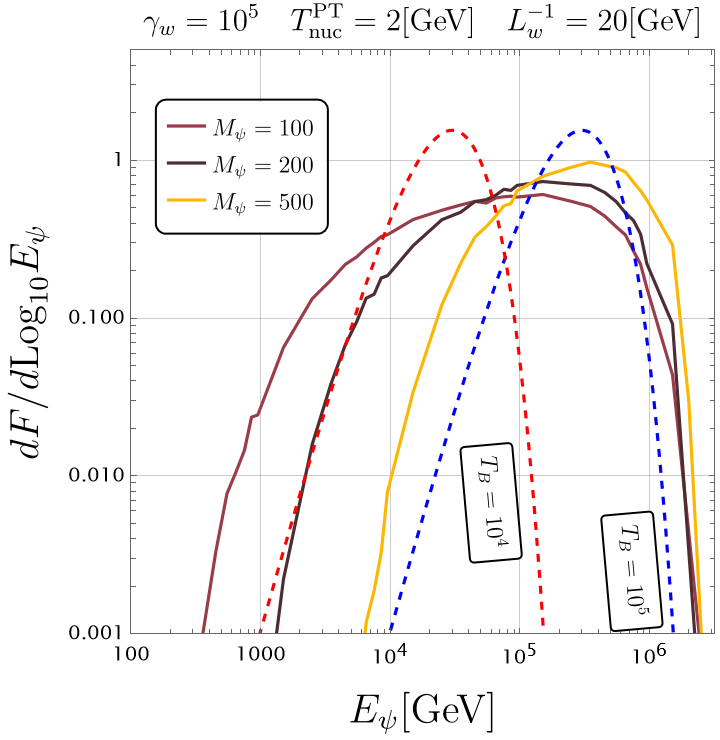}
    \caption{Normalised spectrum of the fermions $\psi$ immediately after emission, via $h \to \psi \psi$. The thick lines represent the numerical solutions for the spectrum of $\psi$ from bubbles, where $E_{\psi}$ is the energy in the \emph{plasma frame}. The dashed lines represent the spectrum from a hypothetical thermal Boltzmann abundance of $\psi$ at high temperature ($T_B=10^4, 10^5$ GeV), which are shown as a comparison for the large $E$ behaviour. }
    \label{Fig:spectrum_num}
\end{figure}

The energy distribution of the DM produced via bubble-plasma collisions will not have \blue{the} thermal shape immediately after the production. This spectrum can be obtained 
numerically by convolution  of the Boltzmann distribution for 
the initial particles $h$ with the probability for the heavy-particle production (see details in Appendix \ref{app:spectrum_at_em}). The results are presented on Fig.~\ref{Fig:spectrum_num}, 
where we have used the  same normalisation
\bea 
\int^{\infty}_0 dE_\psi \frac{dF}{dE_\psi} = 1 \, , 
\eea 
for every curve. Here
$d F/dE_\psi$ corresponds to the differential number density and $E_\psi$ is the energy in the plasma frame. 
At low energies the spectrum 
starts 
to rise for $E_\psi \gtrsim M_\psi$,  follows a 
plateau which is finally exponentially cut off by the Boltzmann suppression at $E_\psi \gtrsim \gamma_w  L^{-1}_w \sim \gamma_w 
v$\footnote{Strictly speaking the  dark matter is produced during the whole process of the bubble wall expansion, during which $\gamma_w$ is growing. However the DM production will be dominated by the bubbles with the maximal radii just before the collision, thus we believe the corrections will be subleading.
}.


\paragraph{CDM region}
On the other hand, if the interactions $\psi h \to \psi h$ are active after the phase transition
\bea 
\label{eq:recoupled}
 \frac{T_{\rm reh} M_{\rm pl}}{ \Lambda^2} > 1.66 g_\star^{1/2}(8\pi^3)\, ,
\eea 
$\psi$ will quickly slow down until it  reaches a kinetic distribution with strongly non-relativistic velocities. DM then becomes usual Cold DM (CDM).

\paragraph{Transition region between CDM and FS.}
We now address the region  between two extreme cases in Eq.\eqref{eq:decoupled} and Eq.\eqref{eq:recoupled}. In order to find the DM velocity $V_{\rm eq}$ in this transition regime, we need to take into account the momentum loss due to rescattering of DM with plasma and follow the average DM energy evolution as a function of $T$. This effect can be estimated from the following simplified procedure.
The momentum of a particle lost in one collision $\psi(p_1) h(p_2) \to \psi(p_3) h(p_4)$, $\delta p_{\psi}$,  is given by  \cite{Baldes:2022oev}
\bea 
\delta p_{\psi} \approx E^\psi_1 - E^\psi_3 \approx -t/4T\, , 
\eea 
where the energies are evaluated in the plasma frame and $t \equiv (p_1 - p_3)^2 \approx -2 p_1 \cdot p_3$ is the usual Mandelstam variable. Now the equation for the evolution of the $\psi$ average energy $\bar E$, as long as $\bar E \approx |p_1|\gg M_\psi$, is 
\bea
\label{eq:evolution_eq}
a^{-1}\frac{d (\bar E a)}{d t}= \vev{\sigma_{h \psi \to h \psi} v_M}\langle \delta  p_\psi \rangle_{\rm loss} n_h \, ,
\eea
where $\langle \delta p_\psi \rangle_{\rm loss} \propto \bar E$ is the average energy loss in one collision and $ v_M$ is the M{\o}ller velocity. Using that $d\sigma_{h \psi \to h \psi}/dt \propto t/s^2\Lambda^2$, the energy of the collision being approximately $s \approx 4ET$, and neglecting the mass $M_\psi$ for simplicity of the computation, we obtain
\begin{align}
   - \langle \delta p_X \rangle_{\rm loss}(E) = -\frac{1}{\sigma_{h \psi \to h \psi}}\int^s dt \frac{d\sigma_{h \psi \to h \psi}}{dt}\delta p_\psi=
   \frac{2 E}{3}  \, . 
\end{align} 
   Then we can rewrite the evolution Eq.\eqref{eq:evolution_eq} as follows
\bea
\frac{d }{ dt}= - H T\frac{d}{d T} \qquad \Rightarrow\qquad 
 a^{-1} H T \frac{d (a \bar E)}{d T}= \vev{\sigma_{h \psi \to h \psi} v_M} \bar E \frac{2 \zeta (3)}{3\pi^2} T^3 \, ,
\eea
where we have assumed that $g_*$, the relativistic number of d.o.f, is not changing and we used that the number of d.o.f. of the scalar $h$, $g_h = 1$. The velocity averaged cross-section for the dimension five operator is given by:  
\bea
\vev{\sigma_{h \psi \to h \psi} v_M }= \frac{1}{8\pi\Lambda^2},
\eea
where we approximated the relative velocity to be $v_M \approx 2$. And we finally obtain the evolution Eq.\eqref{eq:evolution_eq} in the form
\bea
\frac{d (a \bar E)}{a \bar E}\approx \frac{1}{12\pi\Lambda^2} \frac{\zeta(3)M_{\rm pl}}{1.66 \pi^2 \sqrt{g_*} }dT \, .
\eea
The evolution equation  can then be trivially integrated to give the final energy of the $\psi$ particle at the decoupling of the scattering
\bea
\frac{E_f}{E_i}= \l(\frac{T_f}{T_i}\r)  \exp \l[-\frac{1}{12\pi\Lambda^2}\frac{2 \zeta(3)M_{\rm pl}}{1.66\pi^2 \sqrt g_*}(T_i-T_f)\r] \,. 
\eea
In this expression, the initial temperature $T_i$ is the reheating temperature \emph{after the transition} $T_{\rm reh} = T_i$. Since this expression was derived using the assumption of relativistic $\psi$ we can use it till the temperature $ T_f = T_{\rm NR}$ when the $\psi$ becomes non-relativistic and $E_\psi \approx M_\psi$. Then after the end of the interactions, the velocity will be simply redshifted by the universe expansion. The velocity at matter-radiation equality is thus given by
\bea
\label{eq:T_NR_solve}
&&V_{\rm eq} = \bigg(\frac{g_{\star}(T_{\rm eq})}{g_{\star}(T_{\rm NR})}\bigg)^{1/3}
\frac{ T_{\rm eq}}{T_{\rm NR}},
\qquad \frac{M_\psi}{\bar E_{\psi} } \simeq  \frac{T_{\rm NR}}{T_{\rm reh}}  \exp \l[ - \frac{1}{6\pi \Lambda^2} \frac{ \zeta(3)M_{\rm pl}}{1.66\pi^2 \sqrt g_*}(T_{\rm reh}-T_{\rm NR})\r] \, .
\eea
where $\bar E_{\psi} $ is the average energy given in Eq.\eqref{eq:average_E_kperp}. The second relation can be solved for $T_{\rm NR}$ and plugged in the first. Combining this with  $T_{\rm eq} \approx 0.8$ \text{~eV}, we obtain
\bea 
\boxed{V_{\rm eq} \approx 2 \times  10^{-10} \frac{ \text{GeV} \times \bar E_{\psi}}{T_{\rm reh} M_{\psi} }  \exp \l[ - \frac{\zeta(3)}{ 1.66\times 6\pi^3 \sqrt g_*} \frac{ M_{\rm pl} (T_{\rm reh}-T_{\rm NR})}{ \Lambda^2}\r] } \, .
\label{eq: v_eq ferm}
\eea

\paragraph{Dynamics of the phase transition and pressure from production}
We now  sketch the dynamics of the PT. 
Bubbles nucleate generically with a radius 
\bea 
R_{\rm nuc} \propto 1/T_{\rm nuc} \,. 
\eea 
The expansion of a bubble can proceed in two regimes: either the bubble reaches a 
steady state motion and the velocity becomes constant (terminal velocity regime), or 
the bubble keeps accelerating until collision (runaway regime). As long as the pressure from the release of energy 
is not balanced by the pressure from the plasma, the bubble keeps accelerating 
with the equation of motion\cite{PhysRevD.45.3415,Ellis:2019oqb}
\bea
\gamma^{}_w (R)\approx  \frac{2 R}{3 R_{\rm nuc}}\l(1-\frac{\Delta \mathcal{P}}{\Delta V}\r) \approx  \frac{2 R}{3 R_{\rm nuc}} \,.
\eea 
If the bubble keeps accelerating until collision, the largest velocity is controlled by the radius of the bubble at collision, $\gamma_w^{\rm ter} (R_\star)$, given by 
\bea
 R_*\approx \frac{(8 \pi)^{1/3}v_w}{H[T_{\rm nuc}]\beta(T_{\rm nuc})},~~~\beta(T)= T\frac{d}{dT}\l(\frac{S_3}{T}\r) \, , 
\eea
where $R_\star$ is an estimate for the bubble size at collision and $\beta$ the inverse dimensionless duration parameter of the transition. We obtain the boost factor at collision  
\bea 
\label{eq:gamma_coll}
\gamma^{\rm coll
}_w \sim \frac{2 \sqrt{10} M_{\rm pl} T_{\rm nuc}}{ \pi^{2/3}\sqrt{8\pi} \sqrt{g_\star} \beta T_{\rm reh}^2} \approx 0.06\frac{ M_{\rm pl} T_{\rm nuc}}{ \beta T_{\rm reh}^2} \, .
\eea 

However,  the pressure from particles coupling to $h$ might terminate the acceleration long before $\gamma_w^{\rm collision}$ is reached. 
The study of the bubble wall interaction with the plasma is a field under active investigation\cite{Dine:1992wr,Liu:1992tn,Moore:1995ua,Moore:1995si,Dorsch:2018pat,Laurent:2022jrs,Jiang:2022btc,Konstandin:2010dm, BarrosoMancha:2020fay,Balaji:2020yrx, Wang:2022txy,Krajewski:2023clt, Sanchez-Garitaonandia:2023zqz,Bodeker:2009qy, Bodeker:2017cim, Azatov:2020ufh,Gouttenoire:2021kjv,Ai:2023suz,Azatov:2023xem,Ai:2021kak,Ai:2023see, Ai:2024shx, Azatov:2024auq,Barni:2024lkj}. In the regimes of ultra-relativistic bubbles, the computation of the terminal velocities \cite{Bodeker:2009qy, Bodeker:2017cim, Azatov:2020ufh,Gouttenoire:2021kjv,Azatov:2023xem} amounts to comparing the release of energy $\Delta V$ with the plasma pressure in the relativistic regime $\Delta \mathcal{P} (\gamma_w)$. In principle, we could obtain  the terminal velocities by solving \bea \Delta V \approx  \sum \mathcal{P}(\gamma^{\rm ter}_w) \, ,\eea where $\sum \mathcal{P}(\gamma^{\rm ter}_w)$ is the sum of the different source of pressure.

In the relativistic regime, the computation of the pressure however largely simplifies and the following picture emerges: the pressure is due to the interactions inducing an exchange of momentum from the plasma to the bubble wall. Schematically it reads
\bea 
\label{eq:intuitive_picture}
\mathcal{P}^{\gamma_w \to \infty} \approx \sum_{ij}\underbrace{\frac{p_z}{p_0} n_i}_{\text{flux}} \times \underbrace{P_{i \to j}}_{\text{probability $i \to j $}} \times \underbrace{\Delta p_{i \to j}}_{\text{exchange of momentum $i \to j$}}
\eea 
where the first factor is the incoming flux of particle species $i$ entering into the wall and having a transition $i \to j$ , i.e. to state $j$, with an associated loss of momentum $\Delta p_{i \to j} \equiv p_i - p_j$. This loss of momentum of the plasma is transmitted to the wall, which is felt by the wall as a pressure. A more complete presentation is provided in Appendix \ref{app_pressures}.

We now investigate the possibility of having runaway walls, as necessary for our production mechanism. The first contribution to the pressure is the pressure from the particles coupling to the BSM Higgses $h$ gaining mass\cite{Bodeker:2009qy}, 
\begin{align}
\label{eq:mass_gain}
\mathcal{P}_{h} &\approx g_h\frac{m_h^2T^2_{\rm nuc}}{24} \qquad \text{(BSM Higgs gaining mass: model-independent)}\, ,
\\
\mathcal{P}_{i} &\approx c_ig_i\frac{m_i^2T^2_{\rm nuc}}{24} \qquad \text{(Particles coupling to $h$ gaining mass: model-dependent)}\, ,
\\ 
\mathcal{P}_{\rm LO} &=\mathcal{P}_{h} + \sum_i \mathcal{P}_{i} \, ,
\end{align} 
where $m_h$ is the mass of the BSM Higgs in the broken phase, $c_i = 1(1/2)$ for bosons(fermions). We assume that this pressure is not enough to prevent to balance the release of energy $\Delta V> \mathcal{P}_{\rm LO}$. On the other hand, if the BSM Higgs couples sizably with gauge coupling $g$ to gauge bosons, the emission of soft transverse gauge bosons\cite{Bodeker:2017cim, Gouttenoire:2021kjv} and longitudinal gauge bosons\cite{Azatov:2023xem}, would induce a pressure 
\bea 
\label{frictiongauge}
\mathcal{P}_{g} \propto \frac{g^3}{16 \pi^2} \gamma_w T_{\rm nuc}^3 v \qquad \text{(emission of soft bosons: model-dependent)} \, ,
\eea 
preventing runaway. The intuitive picture in Eq.\eqref{eq:intuitive_picture} permits to understand how the pressure on the wall can increase with the energy $\gamma_w T$ without ever threatening unitarity: in the wall frame, $n_i \propto \gamma_w T^3$ while $\Delta p_z \sim v$ and the plasma frame $n_i \propto T^3 $ while $\Delta p_z \sim \gamma_w v$. In both frames, the probability of the emission of the soft gauge boson is bounded $P_{\phi \to \phi A} \ll 1$.  

We thus assume that the BSM Higgs does not couple to gauge bosons so that $\mathcal{P}_{g} \to 0$\cite{Baldes:2024wuz} 
or the gauge coupling is small, e.g., $g\lesssim 0.01\beta^{1/3}\left(\frac{v}{10^{10}\,{\rm GeV}}\right)^{1/3}$ by taking $T_{\rm reh}\sim T_{\rm nuc}\sim (\Delta V)^{1/4}\sim v$ and assuming Eq.\,\eqref{frictiongauge} for the friction, and in this case, the pressure from gauge boson emission remains always subleading.
We finally focus on the unavoidable (model independent) pressure induced by the production. Indeed, the transition $h \to \psi \psi$ has a non-vanishing exchange of momentum, which is transmitted to the wall at the production. Because of this, the wall undergoes a plasma pressure\cite{Azatov:2020ufh} (see Appendix \ref{app_pressures} for the details of the computation)
\bea
\label{eq:pressure_prod}
\Delta\mathcal{P}^{\rm prod}_{h \to \psi \psi}\approx  \frac{1}{8\pi^2}\frac{v^3 n_h \gamma_w}{\Lambda^2}  \,  \qquad\text{(production pressure: model-independent)} \, ,
\eea 
which we call the \emph{production pressure}\cite{Azatov:2020ufh} and could in principle stop the wall acceleration. Let us insist on the fact that the maximal pressure that can be induced in the \emph{context of the EFT validity} is given by 
\bea 
\gamma^{\rm max}_w \approx \frac{\Lambda^2}{2vT_{\rm nuc}} \qquad \Rightarrow \qquad \mathcal{P}^{\rm prod}_{h \to \psi \psi} \bigg|_{\text{ max}} \approx g_h\frac{v^2 T_{\rm nuc}^2}{16\pi^4}  \, ,
\eea 
where we set $2\gamma_w v T_{\rm nuc} = \Lambda^2$. This remains in principle always smaller than the pressure from the $h$ obtaining a mass in Eq.\eqref{eq:mass_gain}. We can thus safely neglect it.

At the end of the day, the boost factor involved in the production of DM is given by
\bea 
\label{eq:gamma_wall}
\gamma^{\rm max}_w = \text{Min}\big[\gamma^{\rm ter}_w, \gamma^{\rm coll}_w\big] \,. 
\eea 

The value of $\gamma^{\rm ter}_w$ strongly relies on the physics of the PT sector, namely the presence of gauge bosons and further particles, as well as the amount of supercooling. Following the discussion above, we however consider the following situation 
\bea 
\Delta V > \mathcal{P}_{h} \gg  \mathcal{P}^{\rm prod}_{h \to \psi \psi} \bigg|_{\text{ max}} , \qquad \mathcal{P}_g \sim 0 \qquad \Rightarrow \qquad \text{Runaway regime}
\eea 
As a consequence, in the remainder of this paper we will always take  $\gamma^{\rm max}_w =  \gamma^{\rm coll}_w$ from Eq.\eqref{eq:gamma_coll}. 

\begin{figure}
    \centering
    \includegraphics[width=.49\linewidth]{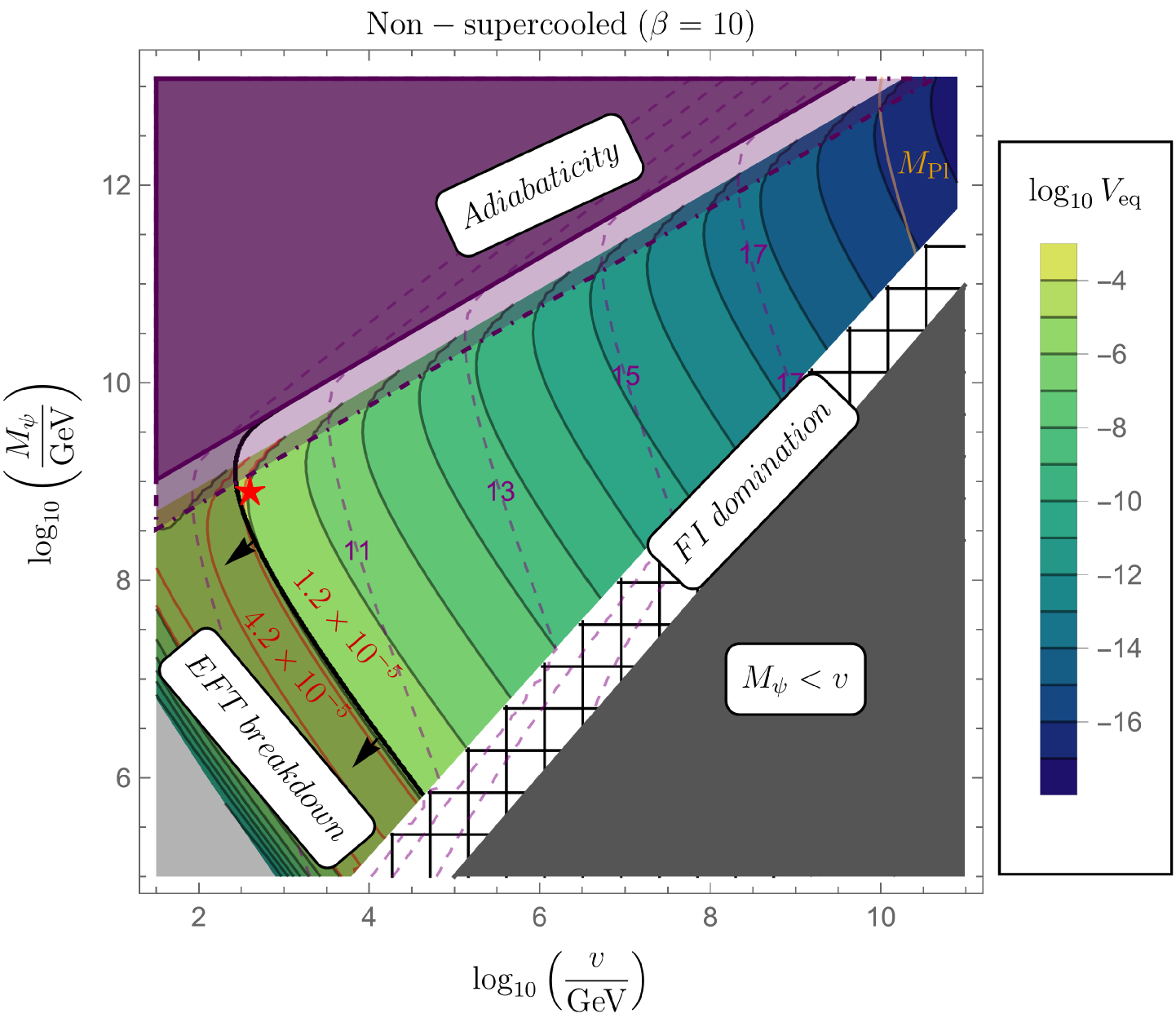}
    \includegraphics[width=.49\linewidth]{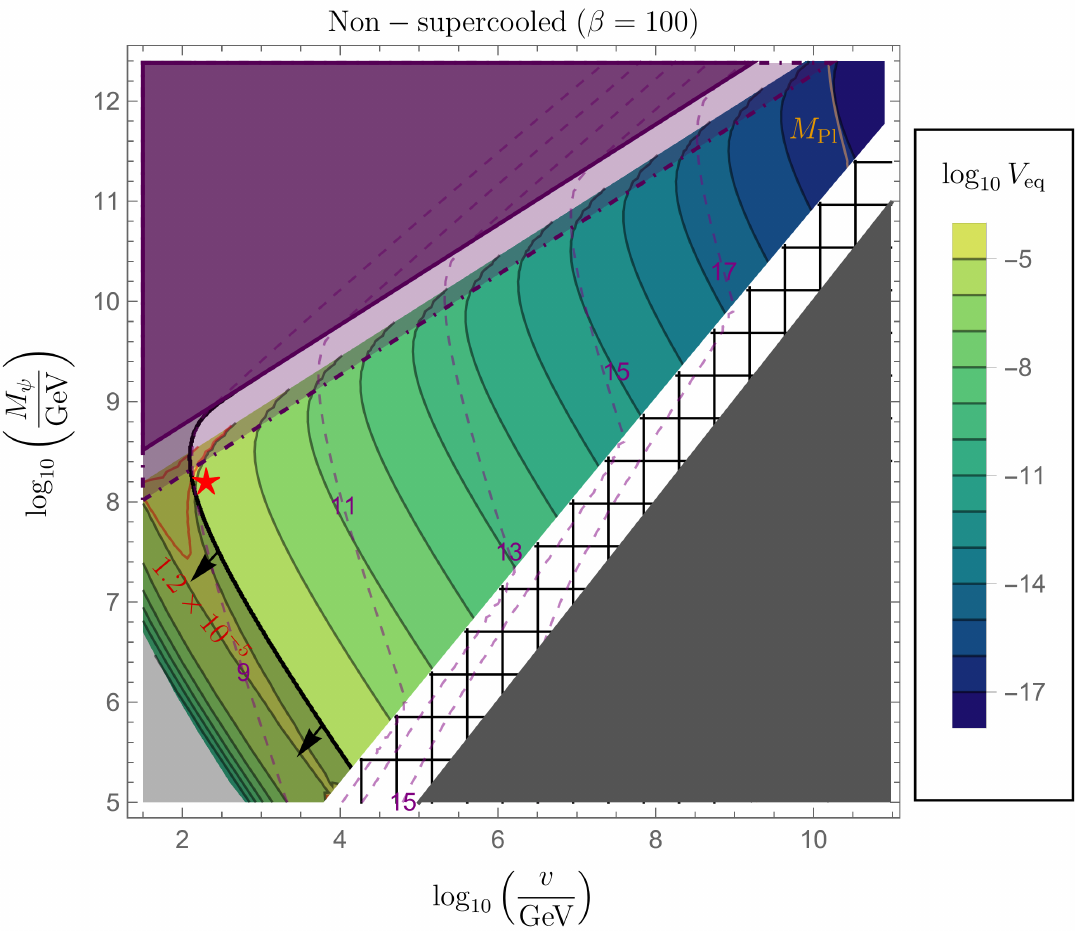}
    \includegraphics[width=.49\linewidth]{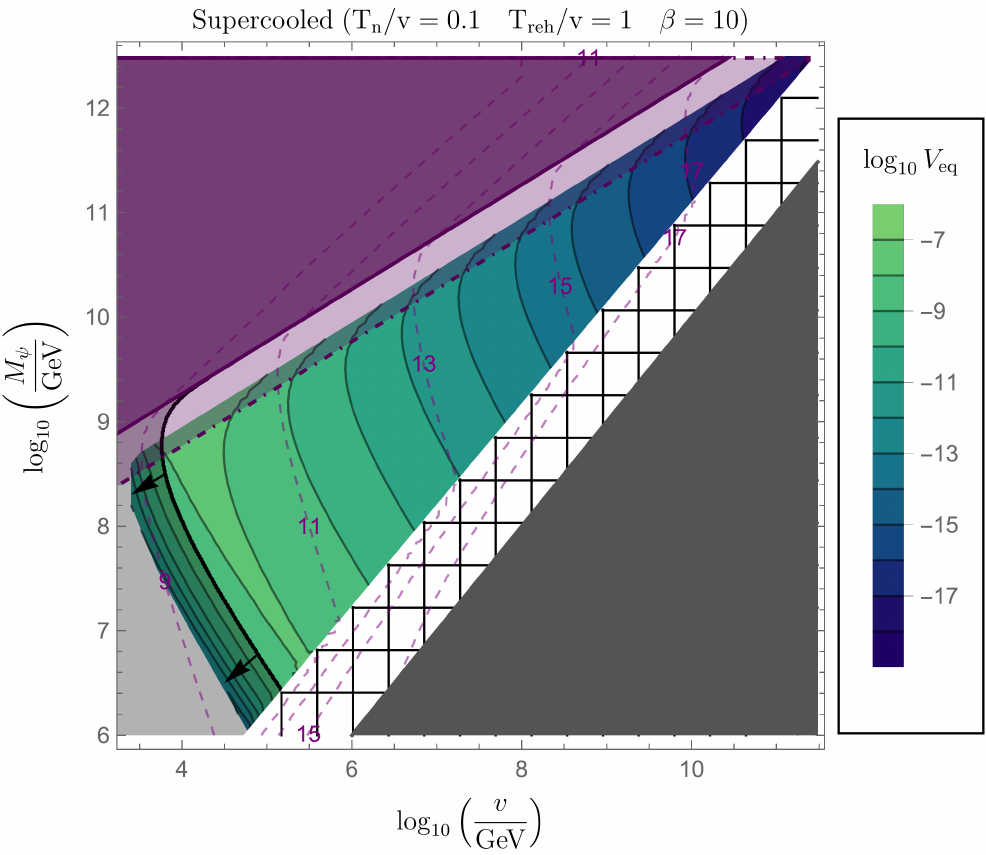}
    \includegraphics[width=.49\linewidth]{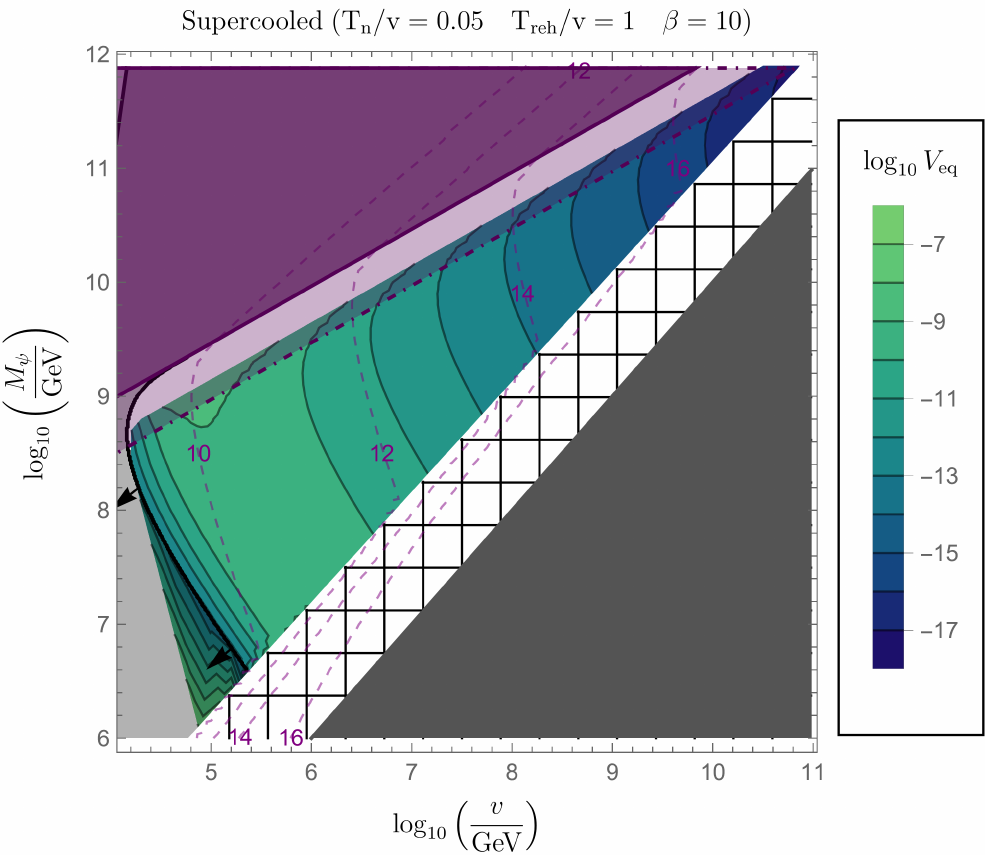}
    
\caption{Contour plots of $\log_{10}(V_{\rm eq})$ for the operator $\frac{h^2\bar{\psi}\psi }{\Lambda}$ and for various values of $\beta$ and $T_{\rm nuc}$, while $T_{\rm reh}=v$.
Purple dashed lines indicate the isocontours of the UV cutoff, 
$\text{log}
_{10}\left(\frac{\Lambda}{\text{GeV}}\right)$.
The shaded area to the left of the  solid black line excludes the region where the EFT 
analysis breaks down, i.e. $2\gamma v T_{\rm nuc} > \Lambda^2$.
The dark gray region in the lower right of each plot indicates $v>M_\psi$. 
The dark purple area indicates the region where the anti-adiabaticity condition 
is not satisfied $\gamma_w T v \leq 2 M_{\psi}^2$. The light purple area denotes the region defined by the conditions  $ M_\psi^2\in\gamma_w T_{\rm nuc} v \times [0.05,0.5]$, in this region DM can be produced, but expression in Eq.(\ref{eq:average_E_kperp}) is not valid.
In the white hatched regions, FI is the dominant process for DM production. The red lines represent the current and future experimental bounds on the velocity, as given by Eqs. (\ref{eq:LyAlphaBoundNow}, \ref{eq:21cmBoundFuture}), and the red star marks the specific points we study in the main text. Light grey area in the lower right corner of the plot indicates the DM in the thermal equilibrium.
}
\label{fig:dim5_money_1}
\end{figure}

\paragraph{Results and discussion}
After describing the dynamics of the phase transition, we now put it all together and study the parameter space.
In principle, we have five parameters in our model 
\bea 
\bigg(\gamma_w, \alpha_{\rm nuc}^{1/4} \approx \frac{T_{\rm nuc}}{v}, v, \Lambda, M_\psi \bigg), 
\eea 
 but for the plots we fix the value of $T_{\rm nuc}/v$ and set $\gamma_w = \gamma_w^{\rm max}=\gamma^{\rm coll}(\beta,T_{\rm nuc},T_{\rm reh})$ to the value at bubble collision \eqref{eq:gamma_coll}.
We take $T_{\rm reh} = v = T_{\rm nuc}$ for the non-supercooled case. The value of $\Lambda$ is obtained by requiring that the produced DM abundance, given by Eq.\eqref{eq:density_today_BE} and Eq.\eqref{eq:density_today_FI};
\bea 
\Omega^{\rm today}_{\psi, \rm today} h^2 = \Omega^{\text{today}}_{\psi,\text{BE}}h^2 + \Omega^{\text{today}}_{\psi,\text{FI}}h^2  ,
\eea 
matches the observation of DM in the universe today. Thus we are left just with two free parameters $v$- scale of the phase transition and $M_\psi$ dark matter mass, which we use for plotting.
In Fig. \ref{fig:dim5_money_1}  we present several versions of this plot for the various values of $\beta$ and $T_{\rm nuc}/v$. 
We observe that the DM can be fairly warm for $v_\phi \in[10^2,10^5]$ GeV, close to the EFT breakdown. We observe an ankle in the $\Lambda$ contours, 
followed by an exponential suppression. 
This change of behaviour can be understood from Eq.\eqref{eq:density_today_BE} where we set $\Omega^{\text{today}}_{\psi,\text{BE}}h^2 = 0.12$ and solve for $\Lambda$. When the DM production becomes exponentially suppressed, the $\Lambda$ drops exponentially to compensate, creating an ankle in the $\Lambda$ contours. 
We note that generically thermal DM is incompatible with EFT (light grey regions in the bottom-left corner of the plot). More specifically, this region was obtained by solving the right  expression of Eq. \eqref{eq:T_NR_solve} for $T_{\rm NR}$ and checking if the DM 
particle is in equilibrium or not at this temperature.  Note that the naive condition of \eqref{eq:recoupled} leads almost to the same contour.

At last we present two benchmark points with very heavy and warm DM:
\begin{itemize}
    \item $\beta= 10$:
\begin{align}
    &v= 400 \text{ GeV}, \quad M_\psi = 8\times 10^8 \text{ GeV}\\ 
    \nonumber
    &\Lambda= 6.3 \times 10^9\text{ GeV}, \quad \gamma_w  = 1.7\times 10^{14}, \quad V_{\rm eq} = 9.5\times 10^{-6} \, .
\end{align}
\item $\beta= 100$:
\begin{align}
    &v= 200 \text{ GeV}, \quad M_\psi = 1.6 \times 10^8\text{ GeV}\\ 
    \nonumber
    &\Lambda= 1.5 \times 10^9\text{ GeV}, \quad \gamma_w  = 3.4\times 10^{13}, \quad V_{\rm eq} = 7.1\times 10^{-6} \, .
\end{align}
\end{itemize}
These points are represented on Fig. \ref{fig:dim5_money_1} by a red star and might be soon observable by structure formation probes like Lyman-$\alpha$ and 21 cm\cite{Sitwell:2013fpa,Munoz:2019hjh} or sub-halo count~\cite{LSSTDarkMatterGroup:2019mwo}. As we will also comment in section \ref{sec:GW_signal}, such points would also likely produce a copious gravitational wave signal that might be detectable with the space interferometer LISA.

\subsection{Production from dimension 6 operator and vector DM}

We can now go through the same steps for the dark vector DM produced via a dimension 6 operator of the form 
\bea 
\frac{h^2 F_{\mu \nu} F^{\mu \nu}}{\Lambda^2} \, ,
\eea 
where $F_{\mu \nu}$ is the field strength of a dark vector $\gamma$. The probability of the production of transverse gauge bosons is again computed in Appendix \ref{App:prodDM} and reads\footnote{Longitudinal mode production, on the other hand, is model dependent. If we consider the mass of the vector obtained from Higgsing, then one can go back to section \ref{sec:reminder} for the production of longitudinal modes by considering that $\phi$ is the BSM Higgs field for the vector's mass. When the mass of the Higgs and the vector are both light, according to the equivalence theorem (whose validity was discussed in \cite{Giudice:2024tcp}), the estimation is the same as the Higgs case. Since the BSM Higgs gets a VEV, it decays but the vector boson does not need to decay because of the unbroken dark charge conjugation symmetry: $\phi\to \phi^*, F\to -F$.}
\bea 
P_{h \to \gamma \gamma}\equiv P_{h \to \gamma^{\pm} \gamma^{\mp}} \approx 
 \frac{v^3p_0 }{\Lambda^4}\frac{ 1}{4\pi^2} \Theta (p_0 v - 2M_\g^2) \, .
\eea
One might be worried that the probability of the interaction increases with the energy and threaten the unitarity when $p_0 \gg \Lambda^4/v^3$. However, we expect the EFT description to break down around $p_0 \sim \Lambda^2/v$, and the UV description to unitarize the theory. 


The abundance produced by bubble expansion is then given by 
\begin{align}
\label{eq:density_BE_photons}
    n_{\g}^{\rm BE, PF} \approx 
    g_h\frac{2\g_w v^3T_{\rm nuc}^4}{(\pi\Lambda)^4}e^{-\frac{ M^2_\g}{T_{\rm nuc} v\g_w}} \quad .
\end{align}
Hence,  with $g_h=1$, the vector DM fraction today is of the form 
\bea
 \Omega^{\rm BE}_{\gamma, \rm today} h^2 \approx {1.1}\times 10^7 \times \bigg(\frac{1}{g_{\star }(T_{\text{reh}})}\bigg) \bigg(\frac{ \gamma_w v^2T_{\rm nuc} M_\gamma}{ \Lambda^4}\bigg)\bigg(\frac{v}{\text{GeV}}\bigg)\bigg(\frac{T_\text{nuc}}{T_{\text{reh}}}\bigg)^3 e^{-  \frac{M_\gamma^2}{vT_\text{nuc} \gamma_w }}   \,. 
\label{eq:relic_ab_gamma}
\eea
As for the fermion production, an unavoidable FI contribution comes from the reheating temperature after the transition, and is given by (see Appendix \ref{App:FI_abundances} for the complete computation)
\begin{align}
\Omega^{\rm FI}_{\gamma, \rm today} h^2 \approx
\begin{cases}
8.76 \times 10^4 \Big(\frac{M_\gamma}{\text{GeV}} \Big)\frac{M_{\rm pl} M^3_\g}{g_\star^{3/2} \Lambda^4} \sqrt{\frac{M_\gamma}{T_{\rm reh}}} e^{-2M_\gamma/T_{\rm reh}}  \qquad \text{when} \qquad  T_{\rm reh} \ll M_\g
    \\
     1.05 \times 10^6 \Big(\frac{M_\gamma}{\text{GeV}} \Big)\frac{M_{\rm pl} T_{\rm reh}^3}{g_\star^{3/2} \Lambda^4}  \qquad \text{when} \qquad  T_{\rm reh} \gg M_\g
\end{cases}
\,. 
 \end{align}

The total abundance is then given by 
\bea 
\Omega^{\rm tot}_{\gamma, \rm today} h^2 = \Omega^{\rm BE}_{\gamma, \rm today} h^2+\Omega^{\rm FI}_{\gamma, \rm today} h^2 \, . 
\eea 
If the mass of the vector boson is below the reheating temperature, $M_\gamma < T_{\rm reh}$, the FI is relativistic, then the ratio of the relic abundances will be
\bea
\frac{\Omega_{\rm BE}}{\Omega_{\rm FI}}\simeq 10^2\l(\frac{ \gamma_w T_{\rm nuc}}{M_{\rm pl}}\r)\l(\frac{v}{T_{\rm reh}}\r)^3\l( \frac{T_{\rm nuc}}{T_{\rm reh}}\r)^3 \approx \frac{6}{\beta}\l( \frac{T_{\rm nuc}}{T_{\rm reh}}\r)^6 \, .
\eea
Here, we have taken into account only FI after the PT and in the second equality we considered the maximal terminal velocity given by Eq.(\ref{eq:gamma_coll}).\footnote{Generically, assuming FI also happened before the immediately after inflation, at $T_{R}$, we have the further contribution
\begin{align}
\Omega^{\rm FI}_{\gamma, \rm today} h^2 \approx 8.76 \times 10^4 \bigg(\frac{M_\gamma}{\text{GeV}} \bigg)\frac{M_{\rm pl} M^3_\g}{g_\star^{3/2} \Lambda^4} \sqrt{\frac{M_\gamma}{T_{\rm R}}}  \bigg(\frac{T_{\rm nuc}}{T_{\rm reh}}\bigg)^3e^{-2M_\gamma/T_{\rm R}} \,, 
 \end{align}
 which again requires 
 \bea 
 M_\gamma \gtrsim 20 T_{\rm R} \, . 
 \eea 
}
With $\beta \sim 10$, we observe that $\frac{\Omega_{\rm BE}}{\Omega_{\rm FI}}$ is most likely to be smaller than one and we hence conclude that in the case of relativistic FI, the BE contribution will be typically subdominant. For this reason we will impose $M_\gamma > v$ in our future computations. 

The conclusion about the pressure from production is the same as in the dimension five case: The pressure on the bubble wall is given by 
\bea 
\mathcal{P}_{h \to \gamma \gamma} \approx n_h \frac{v^4}{\Lambda^4}   \frac{\gamma_w^2 T}{2\pi^2}  \qquad \text{(production pressure)} \, .
\eea 
Following the same reasoning as for the dimension five case, the maximal pressure within the EFT validity regime is
\bea 
\mathcal{P}^{\rm prod}_{h \to \gamma \gamma} \bigg|_{\text{ max}} \approx g_h\frac{v^2 T_{\rm nuc}^2}{32\pi^4}  \, ,
\eea 
where we set $2\gamma_w v T_{\rm nuc} = \Lambda^2$. This remains in principle always smaller than $\Delta V$ and the LO pressure from the $\phi$ obtaining a mass. We can thus safely neglect it. Immediately after the production, the average energy in the plasma frame of the produced vectors is given by (see Appendix  \ref{app:spectrum_at_em} for the numerical and the analytical computation.)
\bea
\bar{E_\g} \approx C_{\gamma}\g_w L_w^{-1}, \qquad C_{\gamma} \approx 0.16 \, . 
\eea

\begin{figure}
    \centering
    \includegraphics[width=.3\linewidth]{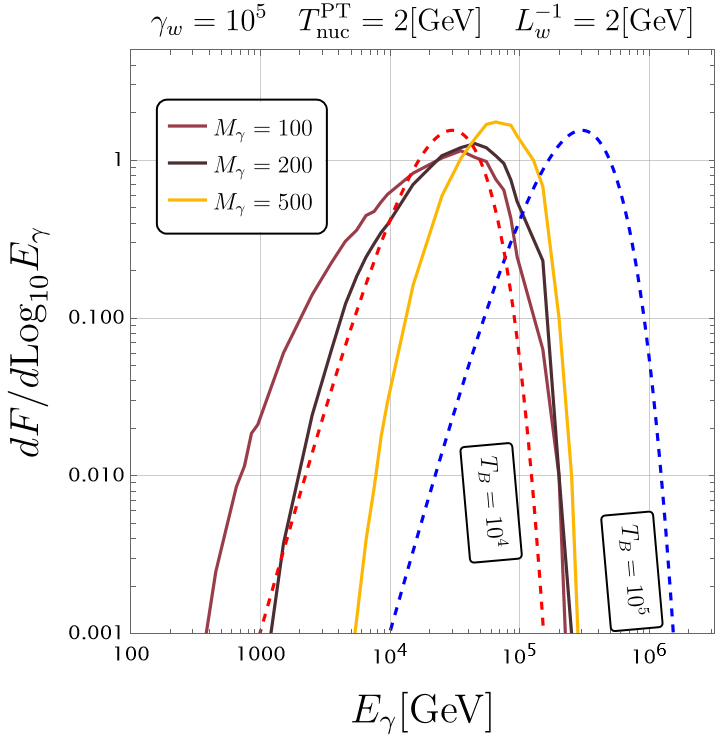}
    \includegraphics[width=.3\linewidth]{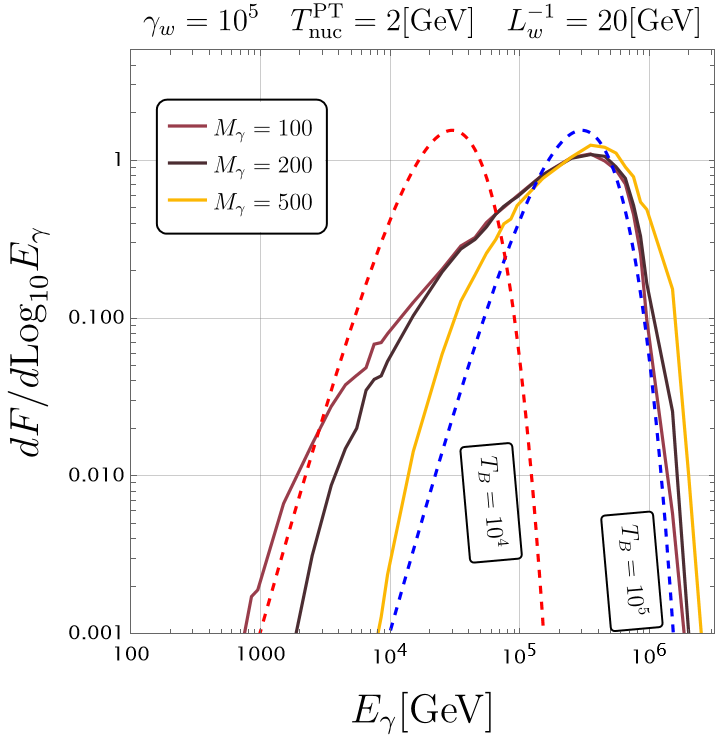}
    \includegraphics[width=.3\linewidth]{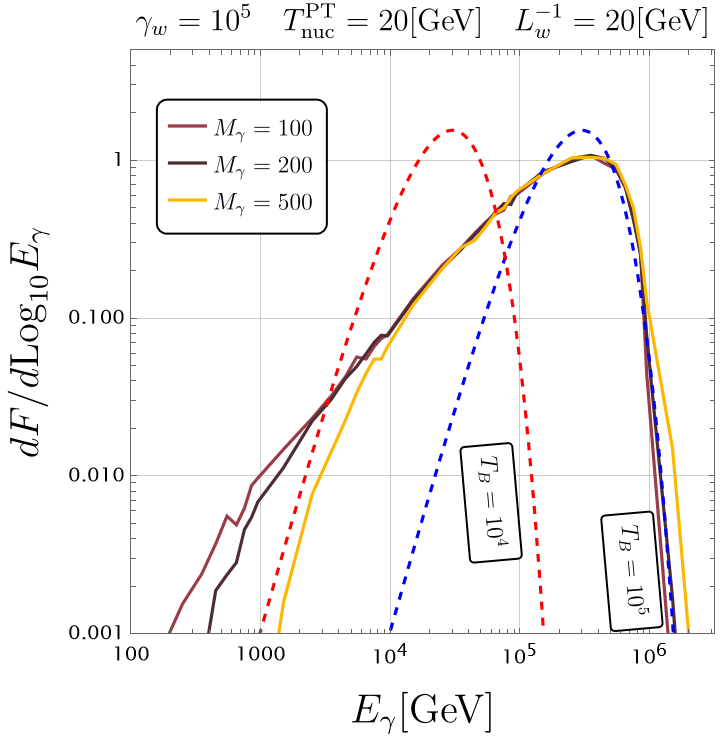}
    \caption{Normalised spectrum of the vector DM $\gamma$ immediately after emission via $h \to \gamma \gamma$. Same as Fig.\ref{Fig:spectrum_num}, but now for the dimension six operator.  }
    \label{Fig:spectrum_num_vectors}
\end{figure}

As for the fermions, the spectrum of the vectors immediately after production is not a thermal spectrum. We present the details of the computation of the spectrum in Appendix \ref{app:spectrum_at_em} and show it for some values of the parameters in Fig.\ref{Fig:spectrum_num_vectors}. The spectrum, compared to the spectrum of the dimension five  case presented in Fig.\ref{Fig:spectrum_num}, does not show a plateau, but an increasing slope until the Boltzmann exponential tail.  As for the dimension five, at low energies the spectrum starts to rise for $E_\g \gtrsim M_\g$.

As before, after production, the vector DM can rescatter via $\gamma h \to \gamma h$, inducing an energy loss that we can compute. The only difference is the energy dependence of the cross-section $\vev{\sigma v}\propto \frac{ T \bar E}{\Lambda^4}$:
\bea
\frac{d (a\bar  E)}{a\bar E^2}= \frac{\zeta(3)}{16\pi^3 1.66\sqrt{g_\star}}\frac{ M_{\rm pl}}{\Lambda^4} T dT  \,. 
\eea
 Then solving for the energy we get the following relations
\bea
V_{\rm eq}=\bigg(\frac{g_{\star}(T_{\rm eq})}{g_{\star}(T_{\rm reh})}\bigg)^{1/3}\frac{ T_{\rm eq}}{T_{\rm NR}}
\qquad M_\gamma \simeq  \frac{T_{\rm NR}}{T_{\rm reh}} C_\gamma\gamma_w v \l[ 1 +\frac{\zeta(3)}{16\pi^3 1.66\sqrt{g_\star}}\frac{ M_{\rm pl} \gamma_w v T_{\rm reh}^2}{\Lambda^4 }\r]^{-1}  \, ,
\eea
which combines to the following expression for the velocity at equality
\bea
\label{eq: v_eq photon}
V_{\rm eq} = \bigg(\frac{g_{\star}(T_{\rm eq})}{g_{\star}(T_{\rm reh})}\bigg)^{1/3} \frac{ T_{\rm eq}}{M_\gamma} \frac{ C_\g\gamma_w v}{T_{\rm reh}} \l[ 1 +\frac{\zeta(3)}{16\pi^3 1.66\sqrt{g_\star}}\frac{ M_{\rm pl} \gamma_w v T_{\rm reh}^2}{\Lambda^4 }\r]^{-1} \, ,
\eea 
giving
\bea 
\boxed{
V_{\rm eq}\approx 
0.32 \times  10^{-10} \frac{ \text{GeV} \times \gamma_w v}{T_{\rm reh} M_{\gamma} }   \l[ 1 +\frac{\zeta(3)}{16\pi^3 1.66\sqrt{g_\star}}\frac{ M_{\rm pl} \gamma_w v T_{\rm reh}^2}{\Lambda^4 }\r]^{-1}
} \, . 
\eea  

\begin{figure}
    \centering
    \includegraphics[width=.49\linewidth]{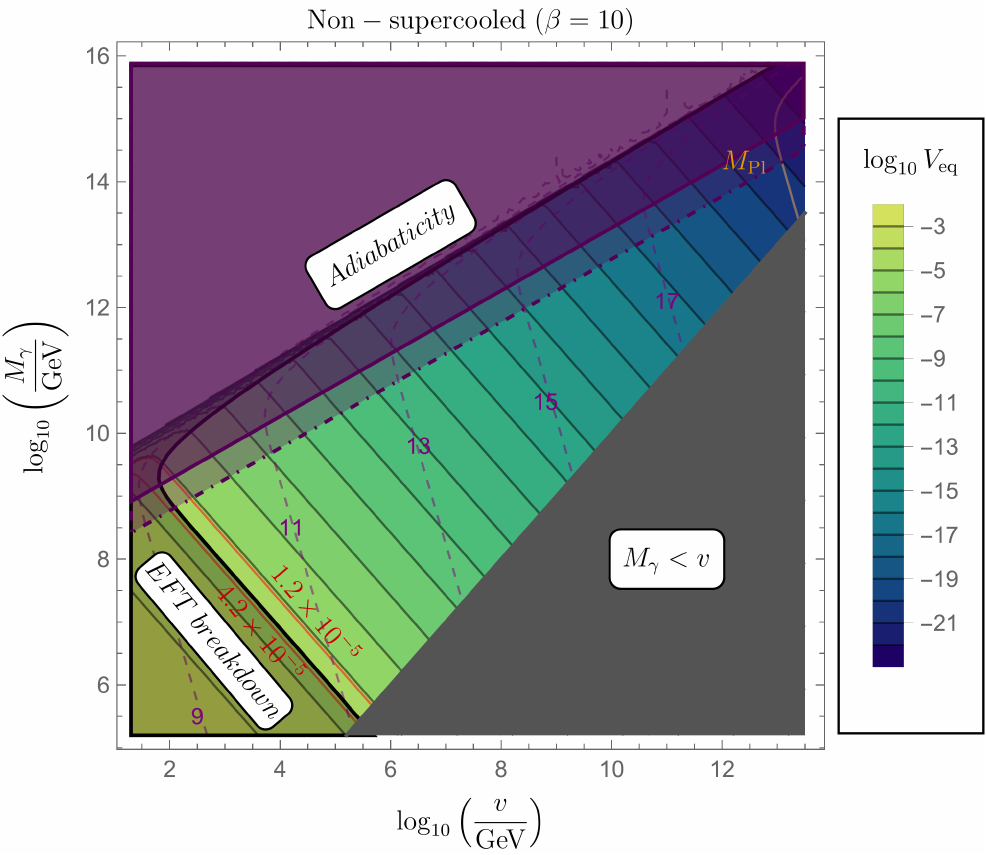}
    \includegraphics[width=.49\linewidth]{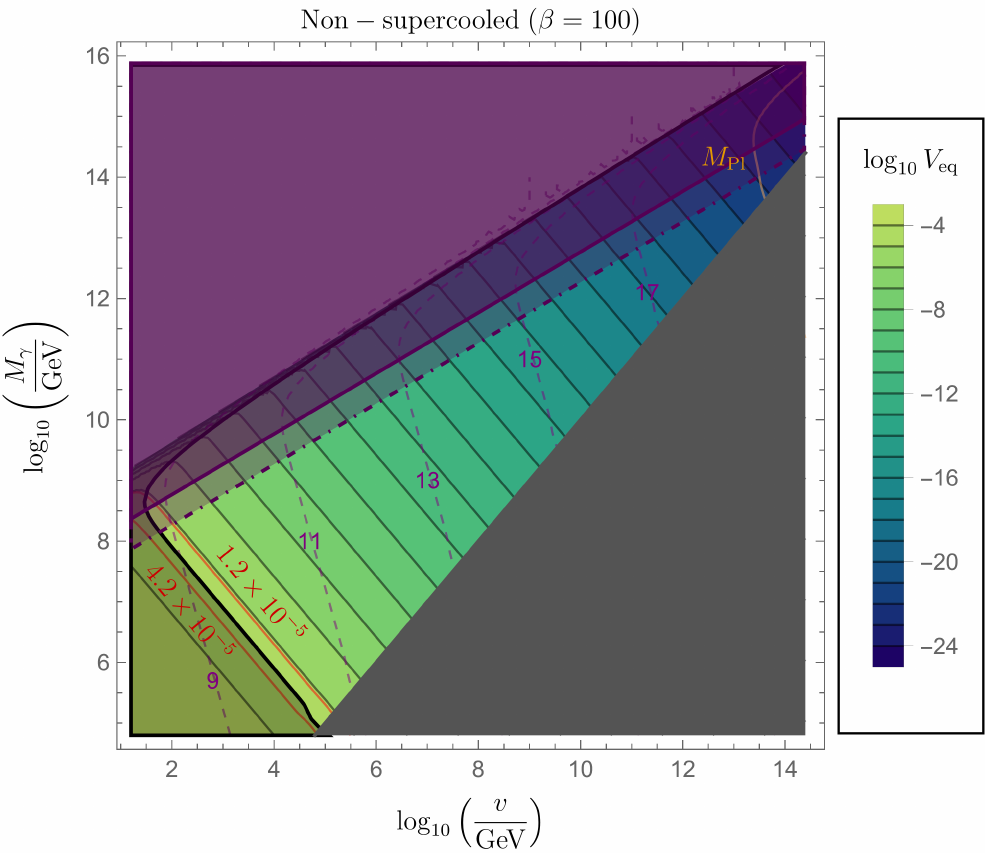}
    \includegraphics[width=.49\linewidth]{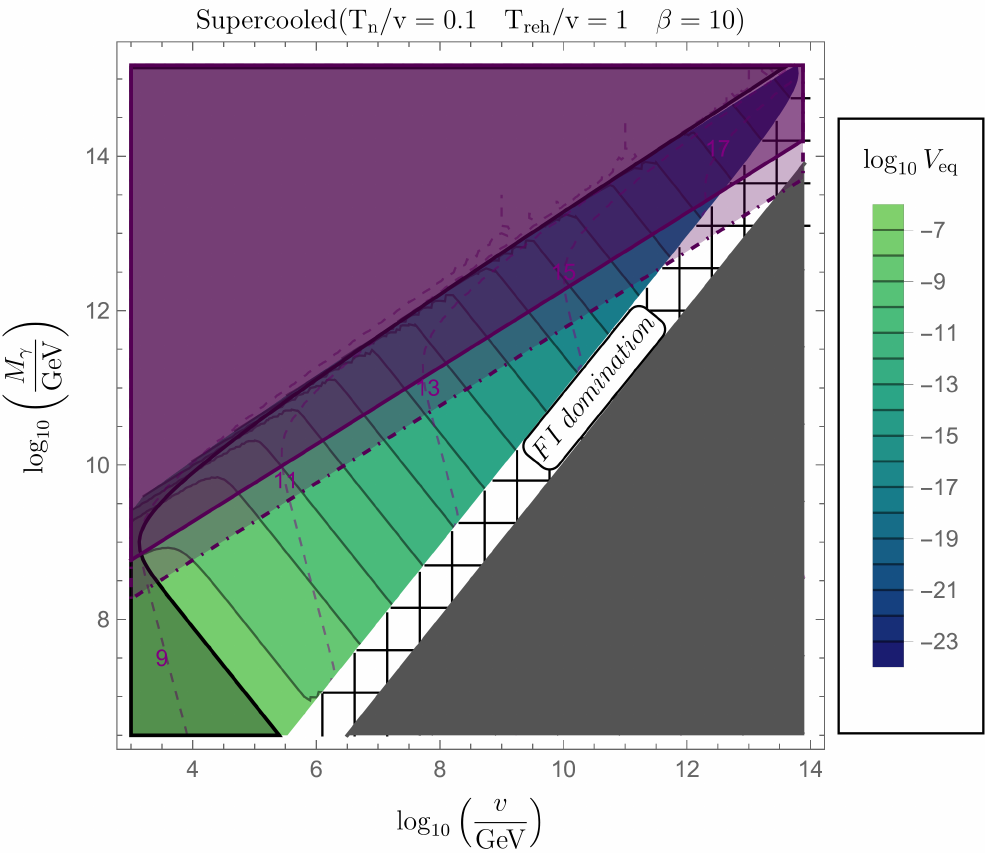}
    \includegraphics[width=.49\linewidth]{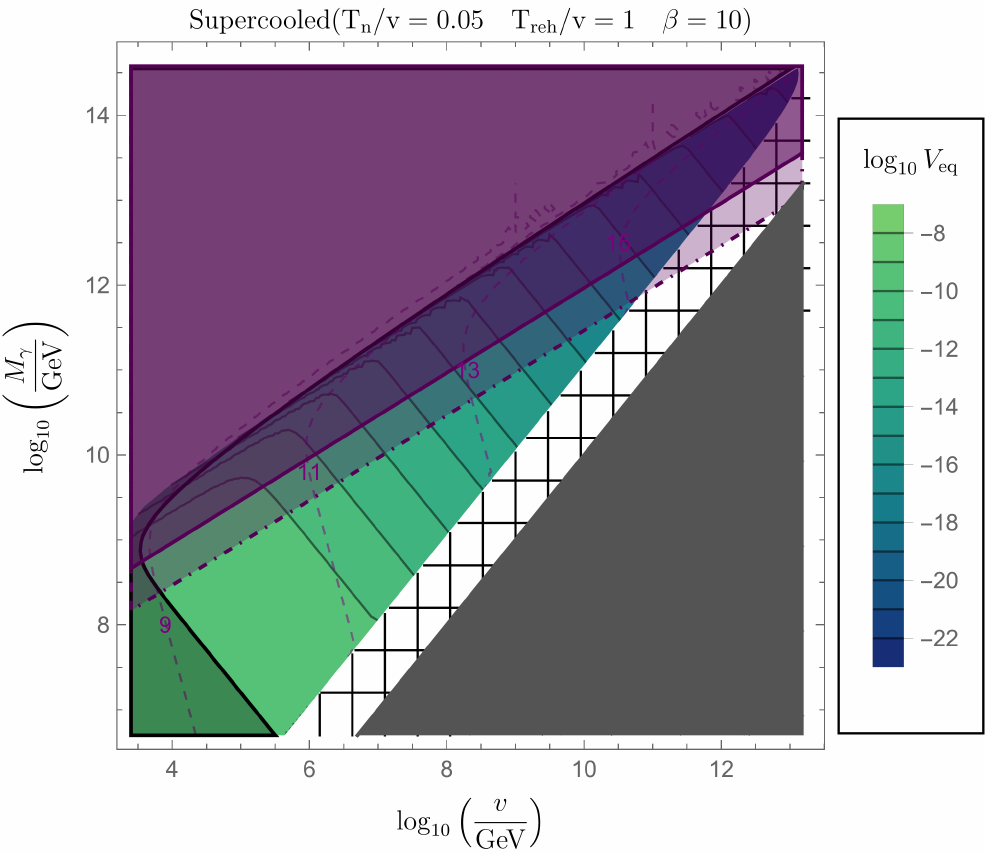}
    \caption{Contour plots of $\log_{10}[V_{\rm eq}]$ for the operator $\frac{h^2 FF}{\Lambda^2}$ for 
different $\beta$, $T_{\rm nuc}$ and $T_{\rm reh}\sim v $. 
Coloring scheme and various contours are exactly the same as in the Fig.\ref{fig:dim5_money_1}
\label{fig:velocity_coun_plot_6_photon}}
\end{figure}

We present results on the Fig.\,\ref{fig:velocity_coun_plot_6_photon},  following exactly the same conventions of the Fig. \ref{fig:dim5_money_1}. The cut-off scale is fixed in order to reproduced the DM abundance and the boost factor $\gamma_w$ is fixed to be the one at collision \eqref{eq:gamma_coll}. The color scheme for the plots is exactly the same as in Fig. \ref{fig:dim5_money_1}.
We observe that we can get WDM with velocities  $V_{\rm eq} \sim (0.5-1)\times 10^{-4}$. 

We now discuss the differences and similarities between the fermion and the vector production.

\subsection{Comparison between the dimension five and the dimension six production}
Let us now comment on the differences between the production of fermions and that of vector particles. 
First, looking at Fig. \ref{Fig:spectrum_num} and Fig. \ref{Fig:spectrum_num_vectors}, we observe that the initial spectrum have different energy dependences:
\bea 
\text{(fermion):} \quad \frac{dF_\psi}{dE_\psi} \propto E^0_\psi, \qquad \text{(vector):} \quad \frac{dF_\gamma}{dE_\gamma} \propto E^1_\gamma \, , 
\eea 
in between the Boltzmann suppressed regime and the rise at small $E$.

Secondly, in both the fermion and the vector case, the EFT validity constraint, requires that the DM is generically \emph{free-streaming} after the production and does not thermalise efficiently. Interestingly we find that the parameter space with $V_{\rm eq}\sim 10^{-5}$ is much larger in dimension six models compared to the dimension five.
For the vector emission, the DM is warm in a larger band near the EFT breaking bound. 
A third observation is that there is no striking difference in the parameter space ($\Lambda, M$) allowing the dark sector to 
contain the total amount of  observed DM. This is because the ratio of the production of vector over fermions 
scales like 
\bea 
\frac{n^{\gamma}_{\rm BE}}{n^\psi_{\rm BE}} \propto \mathcal{O}(0.1)\frac{\gamma_w^{\rm coll} T_{\rm nuc} v}{\Lambda^2} \, ,
\eea
up to logarithmic corrections. The consequence is that the very large $\gamma_w^{\rm coll}$ partially cancel the large suppression from $T_{\rm nuc} v/\Lambda^2$. A similar cancellation does not occur for the FI, where we have
\bea
\frac{n^{\gamma}_{\rm FI}}{n^\psi_{\rm FI}} \propto \frac{M_{\rm DM}^2}{\Lambda^2} \, ,
\eea
and, as we consider $M_{\rm DM}\ll \Lambda$, the FI contribution is less relevant. As a consequence, for both cases, a phase transition at the EW scale \emph{can} produce the total amount of observed DM and than this DM will be typically warm. In this case, $h$ may be a BSM Higgs field relevant to the electroweak phase transition. 

Finally, we find that in both cases we can match the observed DM abundance by taking $\Lambda \sim M_{\rm pl}$,  $v=10^{10-14}$GeV. This is a particularly interesting observation because i) we expect on general grounds any dark sector to be coupled to the thermal sector by at least Planck-suppressed operators, ii) The PT of the scale of $v=10^{10-14}$GeV agrees with the seesaw and Peccei-Quinn scales and happens in various well-motivated models for the neutrino mass or the strong CP problem~\cite{Minkowski:1977sc,Yanagida:1979as,Ramond:1979py,Gell-Mann:1979vob,Mohapatra:1979ia,Shifman:1979if,Dine:1981rt,Zhitnitsky:1980tq}. In those models, ultra-relativistic bubble walls can populate the dark sector particles coupled to the visible sector solely via gravity, and explain the DM abundance. In this scenario, the GW signal features a peak frequency of $10^{2-6}$Hz in addition to the scale-invariant one from cosmic strings, since the models spontaneously break $U(1)$ global/gauge symmetry.


\section{Production of strongly coupled gluons}
\label{sec:prod_strongly}

In this section, we study the case in which the dark sector is a strongly coupled pure Yang-Mills theory, which means that the considered dark sector is purely made of gluons in the deconfined phase or glueballs in the confined phase (for the review see \cite{Carlson:1992fn,Hochberg:2014dra,Forestell:2016qhc}). We will thus explore the phenomenology of bubble wall produced gluons which later hadronize and account for the DM in the universe.

\subsection{Glueballs as WIMP dark matter}
If we assume that the dark sector is controlled by a pure Yang-Mills theory,
the glueball DM does not behave as a WIMP, which annihilates to reduce the comoving number efficiently. The difference is that, since a pion does not exist in the pure Yang-Mills theory, the glueball is the lightest particle of the DS. The glueballs decrease their number density via the $3\to 2$ process of the glueball self-interaction, while the glueballs in the final state are more energetic, transforming the number density into thermal energy. This makes the scenario different from usual freeze-out. On top of this, glueball DM has several interesting properties: 
first, the glueball can only have higher dimensional interactions with other particles that are not charged under $SU(N)$. As a consequence, for glueball DM, the stability can be easily satisfied without imposing any further symmetry.  Second, glueballs have an irreducible self-interaction rate controlled by the confinement scale. This self-interaction may be relevant to the small-scale problem or, more recently called galaxy diversity problem, stating that the rotation curves for spiral galaxies have inner slopes in a diverse range\cite{Vogelsberger:2012ku, Vogelsberger:2014pda,Zentner:2022xux}. Very interestingly, the self-interactions of glueballs might also make the SMBH \cite{Pollack:2014rja} grow faster.

\subsection{The set-up}

We now turn to the mechanism for the population of the dark glueball sector. Assuming a vanishing initial glueball abundance, again, there will be two mechanisms that lead to the population of the dark sector, being FI and BE. Both processes will occur via the following effective operator that couples the strong dark sector $G^{\mu\nu}$ with the  thermal sector $h$, which could be the Higgs or another scalar field in the thermal bath that undergoes the phase transition,
\bea 
\frac{h^2 G_{\mu \nu}G^{\mu \nu}}{\Lambda^2} \,. 
\eea 
This effective coupling will produce gluons via the scattering $hh \to gg$ for the FI production case and via the splitting $h \to gg$ by bubble expansion.\footnote{The gravitational freeze-in production via a purely Planck-suppressed operator was presented in \cite{Redi:2020ffc}.} 

For the FI, as well as for the BE, the strongly coupled sector can be populated in two different ways. The production mechanism can directly produce thermalised gluons or glueballs. Moreover, as the glueballs are not naturally protected by a symmetry, they will have a natural decay channel to a pair of $h$ via the dimension six operator. In this case, the decay rate is too large to explain the dark matter stability. To avoid this fast decay,\footnote{An alternative possibility is to consider a parity conserving dark sector, i.e., the strong CP phase is zero. Then the lightest parity odd glueball gets more stabilized and can behave as dark matter.
This also predicts decaying DM because the parity symmetry is explicitly broken in the standard model sector, and the decay rate of the parity odd glueball is not absolutely zero but suppressed compared with the parity even one. We do not consider this possibility in this paper, because the parity odd one is usually not the lightest dark particle, and it would annihilate into the parity even ones. The annihilation channels also exist, making the discussion more complicated. }
we require that $m_h
 \sim v> M_G \sim 5\Lambda_{\rm conf}$ \cite{Curtin:2022tou}, so that the glueballs cannot directly decay to $h$.  Other decay products involve the SM particles and arise from dimension six operators, and we will discuss this possibility later in this section.

\subsection{Comparison of the FI and the BE in two regimes}

In this section, we estimate the relative abundance from FI and from bubble production, within the two regimes of interest: first, the deconfined gluon phase is never attained during the evolution. In this situation, \textit{glueballs are generated directly}. As we will see below, we anticipate that the FI contribution predominates over the bubble production in this regime. Secondly, \textit{a gluon plasma} forms after the thermalization of the dark sector. We will discuss the conditions necessary for this transition in the following sections.

\paragraph{1. Glueball regime} 
In this regime, the gluons will hadronize immediately to $N_{\rm part}$ free-streaming glueballs. As a consequence,
the abundance of glueballs immediately \textit{after} FI can be obtained following the same computation as for the vector production in Eq.\eqref{eq:FI_rel} and multiplying by the multiplicity factor $N_{\rm part}$\footnote{The multiplicity factor takes into account large number of hadrons(glueballs) produced and grows with energy. Using  \cite{Ellis:1996mzs}  for pure gluonic theory we find
$
N_{part}\propto \exp\sqrt{\frac{2C_A}{\pi b}\log\frac{s}{\Lambda^2}}\sim \exp\sqrt{\frac{24}{11}\log\frac{s}{\Lambda^2}}
$

} and the color factor $N^2-1$.  We obtain
\bea
Y_G^{\rm FI} \approx \frac{45(N^2-1) M_{\rm pl}}{3.32g_\star^{3/2} \pi^7 } \frac{T_{\rm reh}^3}{\Lambda^4} N_{\rm part}(T_{\rm reh}/\Lambda_{\rm conf}) \: ,
\eea 
where the FI production occurs \emph{after} the phase transition.  For the case of BE production, one obtains
\bea 
Y^{\rm BE}_G \approx (N^2-1)\frac{\g_w v^3T_{\rm nuc}}{\pi^6 g_{\star}\Lambda^4} \bigg(\frac{T_{\rm nuc}}{T_{\rm reh}}\bigg)^3  N_{\rm part}(\gamma_w T_{\rm reh}/\Lambda_{\rm conf}) \; ,
\eea 
 which we extracted from Eq.\eqref{eq:density_BE_photons}, and multiplied again by the multiplicity and color factor. The ratio of those two production mechanisms is given by 
\bea 
\frac{Y^{\rm BE}_G }{Y_G^{\rm FI}} \approx \frac{3.32\pi g_{\star}^{1/2} \gamma_w v^3 T_{\rm nuc}}{45 M_{\rm pl} T_{\rm reh}^3} \bigg(\frac{T_{\rm nuc}}{T_{\rm reh}}\bigg)^3 N_{\rm part}(\gamma_w)\approx 2.4\frac{ \gamma_w  T_{\rm nuc}}{ M_{\rm pl} } \bigg(\frac{T_{\rm nuc}}{T_{\rm reh}}\bigg)^3N_{\rm part}(\gamma_w)  \, ,
\eea 
which is bounded by above
\bea 
\frac{Y^{\rm BE}_G }{Y_G^{\rm FI}} \lesssim \frac{1}{\beta}\bigg(\frac{T_{\rm nuc}}{T_{\rm reh}}\bigg)^5N_{\rm part}(\gamma_w) \, ,
\eea 
where we assumed that $T_{\rm reh}\approx v$ and we substituted the value of the terminal velocity given in Eq.\eqref{eq:gamma_coll}. Since we expect $\beta \gtrsim 10$, we estimate that in the case of direct glueball emission, the FI mechanism is most likely dominating, or at least at the same order of magnitude compared to bubble production. We will thus leave this regime aside in the remainder of the discussion and leave it to further studies. 

\paragraph{2. Gluon regime}
 Interestingly, strongly coupled theories can potentially lead to very different DM scenarios compared to a weakly interacting DS. Namely, if the energy transfer from the phase transition field to the strongly coupled sector proves sufficiently efficient, it leads to the formation of a gluon plasma. For this reason, determining the total energy transfer is of paramount importance in order to ascertain if a gluon plasma will be formed. Moreover, the initial energy density of the strong sector will also control the final DM abundance. This is because the $gg \to  ggg$ interaction transforms the initially high-energetic gluon to multiple lower energetic gluons.
Hence, the final number of glueballs depends on the initial energy, governing the final DM abundance. We will therefore now proceed to present the expressions for the energy density of both mechanisms.

The average energy of an emitted gluon produced by BE is similar to the case of dark vectors and reads:
\begin{align}
    \bar{E_g} \approx C_g\g_w L_w^{-1} \sim C_g \gamma_w v\: .
\end{align}
where $C_g$ is expected to be similar to $C_\gamma$, defined in Appendix \ref{app:spectrum_at_em}, and therefore $C_g \approx C_\gamma \approx 0.16$. 
To obtain the energy density deposited in the gluon sector, the above expression has to be multiplied by the number density given by Eq. \eqref{eq:density_BE_photons}. Moreover, the energy density of the dark sector immediately after FI has been already computed in Appendix \ref{App:FI_abundances}. The energy densities by BE and FI are hence given by 
\bea  
\rho_{\rm BE} \sim n_g \bar E_g \approx (N^2 -1)C_g\frac{2\g^2_w v^4T_{\rm nuc}^4}{(\pi\Lambda)^4}  \, ,\qquad \rho_{\rm FI} \approx  10.2 \frac{(N^2-1)T^7_{\rm reh}M_{\rm Pl}}{ \sqrt{g_\star}\pi^5 \Lambda^4} \: ,
\eea
respectively. Notice the quadratic dependence on the boost factor $\gamma_w$ for the case of BE production. This can be understood in the following way: the energy density is given by $E_g n_g$  where both quantities, $n_g$ and $\bar E_g$ are Lorentz boosted and scale like $\gamma_w \gg 1$. Of course, similar to the weakly coupled cases, a condition for the validity of our computation of the BE-produced abundance is that the EFT is still valid for large values of $\gamma_w$ that are considered, i.e.
\bea 
\label{eq:EFT_cond}
\Lambda^2 > s \approx  2\gamma_w v T_{\rm nuc}  \qquad \qquad \text{(breakdown of the EFT)} \,.
\eea 

These expressions allow for a straightforward comparison
\bea 
\label{eq:ratio}
\frac{\rho_{\rm BE}}{\rho_{\rm FI}} \approx C_g\frac{ 2\pi \gamma_w^2 v^4 T_{\rm nuc}^4}{T_{\rm reh}^7 M_{\rm pl}} \sim 2\pi C_g \frac{\gamma_w^2 T_{\rm nuc}^4}{v^3 M_{\rm pl}}  \sim 0.02C_g\frac{M_{\rm pl}}{\beta^2 v} \bigg(\frac{T_{\rm nuc}}{v} \bigg)^6 \: ,
\eea 
where we used Eq. \eqref{eq:gamma_coll} in the last step. We notice a very large enhancement due to the factor $M_{\rm pl}/v$, which makes the BE production largely dominant if the cooling is mild. We however notice that if the cooling is strong $T_{\rm nuc} \ll 10^{-2} v$, then it is unlikely that the BE mechanism can dominate.
As the case of mild cooling appears to be more interesting phenomenologically, we will focus on it from now on. 

 Lastly, we would like to comment on the glueball production in the bubble-bubble collisions, similar to the processes described in Ref.\cite{ Falkowski:2012fb,Mansour:2023fwj,Giudice:2024tcp,Shakya:2023kjf }. Unlike in weakly coupled theories, energy transfer 
to the dark sector will be more efficient, as the  glueballs will have energy $\sim \gamma v$. However, the parameter space where EFT description is  valid is much smaller for the bubble-bubble 
collisions compared to plasma-bubble case. That said, the importance of the bubble-bubble collisions remains an open question and 
we leave the analysis of this process for the future studies.

\subsection{Conditions for the formation of a gluon plasma}
\label{section: CondGluonPlasma}
We have observed that if the dark sector reaches a gluon plasma, the BE mechanism is likely to dominate over the associated FI contribution for non-supercooled PTs. We can wonder now: what are the conditions to reach such gluon plasma state after the phase transition?  We find two conditions: 

\bit
\item First, one should transfer sufficient energy to the dark sector
\bea 
\label{eq:qg_enough_E}
\rho_g  \approx  C_g (N^2 -1)\frac{2\g^2_w v^4T_{\rm nuc}^4}{(\pi\Lambda)^4} >\Lambda_{\rm conf}^4 \, .
\eea 
    \item Second, even if the energy transfer to the dark sector is large enough, but the initial density of glueballs is very small, a gluon plasma will not be formed. To estimate the minimal necessary density, we require that the scattering rate of glueballs is larger than the Hubble expansion rate:
\bea 
\label{eq:GG_from_BE}
\Gamma_{G G \to GG}  \sim \Gamma_{GGG ...}   > H \qquad \text{(Thermalisation of the dark sector)} \; ,
\eea 
where the rate $\Gamma_{GGG ...}$ is the total  rate of the processes in which three or more glueballs are produced in $GG$ collisions. It  would scale parametrically the same as $\Gamma_{GG\to GG}$. We can estimate this rate as follows: 
\bea
\Gamma^{\rm BE}_{GGG ...} \sim n^{\rm BE}_G\times \sigma_{GGG ...}\approx \sigma_{GGG ...}\times C_g(N^2-1)\frac{2\g_w v^3T_{\rm nuc}^4}{(\pi \Lambda)^4} N_{\rm part}\bigg(\frac{\gamma_w v}{\Lambda_{\rm conf}}\bigg)   
\eea
where we have used Eq.\eqref{eq:density_BE_photons}
    $n_{G} \approx N_{\rm part} n_\gamma \approx 
    N_{\rm part}\frac{2\g_w v^3T_{\rm nuc}^4}{(\pi \Lambda)^4} $.
The scattering amongst glueballs can be estimated by naive dimensional analysis. 
The estimate for the cross section of $2\to 3$ processes then reads \cite{Forestell:2016qhc} 
\bea
\sigma_{GGG ...}\sim \sigma_{GG\to GG}\sim \frac{(4\pi)^3}{N^2\Lambda^2_{\rm conf}}\sim \frac{220}{\Lambda_{\rm conf}^2}\l(\frac{3}{N}\r)^2 \, ,
\eea
which is similar to the observed total QCD cross section $\sigma_{\rm QCD}\sim \frac{250}{ GeV^2}$ \cite{TOTEM:2017asr}.
If this condition is fulfilled, the interaction $GG\to GGG$ (or more glueballs in the final state) is active and the number density of glueballs starts to grow very quickly leading to the formation of the 
gluon plasma. Note that the composite glueball has a cross-section of $\sim 1/\Lambda^2_{\rm conf}$ even at the high energy collision, similarly to the nucleon scattering.\\
 
So far, we only considered collisions between fast glueballs produced via BE. However a fast glueball produced via BE could have a collision with a slow glueball produced via FI. In this setting, there are always two populations of glueballs: 1) some cold and slow glueballs coming from FI and 2) some very boosted glueballs produced by the bubble. Population 1) will naturally serve as targets for the fast and scarce glueballs of population 2) to induce $GG \to GG...$ interactions. If this process is efficient it will also lead to the rapid increase of the glueball density and the formation of the quark gluon plasma. The scattering rate can be estimated as
\bea
\label{eq:GG_from_FI}
\Gamma^{\rm FI}_{GGG ...} \sim n^{\rm FI}_G \sigma_{GGG ...}   \approx 220\l(\frac{3}{N}\r)^2  \frac{2(N^2-1) }{3.32g_\star^{1/2} \pi^5 } \frac{T_{\rm reh}^6 M_{\rm pl}}{\Lambda^4 \Lambda^2_{\rm conf}} \approx 0.4 \frac{v^6 M_{\rm pl}}{\Lambda^4 \Lambda^2_{\rm conf}} . 
\eea

In what follows, we will take 
\bea 
\label{eq:GG_active}
\Gamma_{ GGG ...} = \Gamma^{\rm FI}_{GGG ...} + \Gamma^{\rm BE}_{GGG ...} \,  > H.
\eea 
as a sufficient condition for the rapid increase of the gluon number density. Thus, we will use the conditions in Eq.\eqref{eq:qg_enough_E} and Eq.\eqref{eq:GG_active} as a criteria for the dark sector being in the deconfined phase. This naive picture will be modified if the following processes, 
possibly relevant for the thermalisation, are active $h G \to h G$,  $h G \to h GG$ and $GG \to hh$. The rate of those interactions is roughly
\begin{align}
\label{eq:rate_phi_g}
\Gamma_{h G\to h GG} &\sim \Gamma_{h G\to h G}  \approx  \frac{n_h \bar E_g T_{\rm reh}}{ 8\pi \Lambda^4} \sim  C_g\frac{\gamma_w T_{\rm reh}^4 v}{8\pi^3 \Lambda^4} \, , 
\nonumber \\
\Gamma_{G G\to hh} &\sim n_G \frac{\gamma_w v}{8\pi \Lambda^4} \, .
\end{align}

However those effects are only relevant if only
\bea 
H < \Gamma_{G G\to hh} \qquad H < \Gamma_{h G\to h GG},
\eea 
but the viable space of parameters of our mode requires $\Lambda_{\rm conf}\ll v$. As we will confirm below, we find that this effect is never relevant. Moreover, in all the parameter space where $\Gamma_{G G\to hh}$ might be active, the rate for thermalization of the dark sector $\Gamma_{GG \to GGG}$ is much larger and lead to a gluon plasma before energy might be given back to the thermal sector. We can then always neglect it.

\eit
We now turn to the abundance of the glueball DM today.

\subsection{The abundance of Glueball DM today}

We here restrict to the region of the parameter space where the gluon plasma is reached after thermalisation.
In this case, the produced glueballs interact with each other and melt to a thermalized gluon plasma soon after the production around the cosmic temperature $T=T_{\rm reh}$. 
The process is as follows: the 
energetic dark gluons produced by the BE form glueballs because, at production,
gluons are initially underdense compared to $\Lambda^{3}_{\rm conf}$. 
Then, due to scattering, the typical energy of each glueball gradually decreases while the total number of 
glueballs increases. When the number density of glueballs becomes higher than $\sim \Lambda^3_{\rm conf}$, the glueball picture no longer holds and we get a dark gluon plasma. The number density continues to increase (see Ref.\,\cite{Kurkela:2011ti}  for the detailed 
thermalization in the $\Lambda_{\rm conf}\to 0$ limit.) until the system reaches thermal equilibrium. 
We consider that this thermalization 
completes within $O(1)$ Hubble time because of the nature of the strong interaction. Then the whole gluon sector will be characterized just by the corresponding temperature $T_{g}$.
The interaction between these gluons with temperature $T_{g}$ ($\ll T_{\rm SM}$, i.e. the temperature of the SM sector, which is also the temperature for the BSM Higgs $h$) and the thermal sector $h$ is negligible due to 
the nature of the higher dimensional interaction, and we call the gluon 
sector, the dark sector. 
Afterwards, due to the expansion of the universe, $T_{g}$ as well as $T_{\rm SM}$ redshifts. When $T_{g}$ becomes smaller than $\Lambda_{\rm conf}$, the glueballs are formed again due to the confinement phase transition, for which we assume that the entropy production is negligible.

The calculation of the glueball relic abundance proceeds in a standard way \cite{Carlson:1992fn,Hochberg:2014dra,Forestell:2016qhc} and in our discussion 
we will follow closely the notations of \cite{Acharya:2017szw}. Then introducing the parameter 
\bea
B\equiv \frac{T_{g}^4}{T_{\rm SM}^4}= \frac{g_{\star} \rho_{g}}{2(N^2-1)\rho_{\rm SM}} = \frac{ 30\rho_{g}}{2\pi^2(N^2-1)T_{\rm SM}^4} =\frac{30C_g}{\pi^6}\frac{ \gamma_w^2  v^4}{ \Lambda^4}\bigg(\frac{T_{\rm nuc}}{T_{\rm reh}}\bigg)^{4} \, ,
\eea
so that the relic abundance of glueball DM today is given by
\bea
&&\frac{(\Omega h^2)_G}{(\Omega h^2)_{\rm DM}}\approx 0.056 (N^2-1)\l(\frac{B}{10^{-12}}\r)^{3/4}
\l(\frac{\Lambda_{\rm conf}}{\rm GeV}\r)W\l[2.1\frac{N^2-1}{N^{18/5}} B^{3/10}\l(\frac{M_{\rm pl}}{\Lambda_{\rm conf}}\r)^{3/5}\r]^{-1} \; ,
\eea
where the function $W$ is the inverse function of $x e^x$.
The value of $\Lambda$ allowing to match the observed abundance of DM can be approximately found by taking the $W \sim 1$, and we obtain
\bea
\label{eq:Lambda_est}
\Lambda_{\text{observed DM}} \approx 10 \bigg(\frac{M_G}{\text{GeV}}\bigg)^{1/3}\bigg(\frac{vT_{\rm nuc}}{T_{\rm reh}^2}\bigg) \bigg(\frac{M_{\rm pl}T_{\rm nuc}}{\beta}\bigg)^{1/2} \to 10 \bigg(\frac{M_G}{\text{GeV}}\bigg)^{1/3} \bigg(\frac{M_{\rm pl}v}{\beta}\bigg)^{1/2}\,,
\eea 
where the last limit is the case of the non-supercooling. We however use the exact expression in the plots.

\subsection{The production during the EWPT}
\label{production in EWPT}
It is particularly interesting to consider the case in which the $h$ field, producing the gluonic sector, is actually the Higgs field. Although this is not the case for the previous vector field and fermion case, the glueballs are produced differently.
In the context of the EWPT, assuming it is a first order PT it has been shown in Ref.\cite{Bodeker:2017cim} that the pressure from the heavy gauge bosons, $W^{\pm}$ and $Z$, on the bubble wall prevents the EWPT from realistically having highly relativistic boost factors. It was shown that $\gamma_w$ scales like $(v/T_{\rm nuc})^3$\cite{Azatov:2022tii}, hence, ultra--relativistic speeds are reached at the price of a long supercooling of the EWPT. 

Using the estimate in Eq.\eqref{eq:ratio}, we find
\bea 
\frac{\rho^{\rm BE}}{\rho^{\rm FI}} \bigg|_{\rm EWPT} \approx 2\pi C_g \frac{\gamma_{EWPT}^2 T_{\rm nuc}^4}{v_{\rm EW}^3 M_{\rm pl}} \propto \frac{v_{\rm EW}}{M_{\rm pl}}  \bigg(\frac{T_{\rm nuc}}{v_{\rm EW}}\bigg)^{-2} \ll 1 \,. 
\eea 
From those estimates, we can conclude that during the EWPT, the production of the dark gluon sector via BE is far too subdominant and can be discarded.

\subsection{Phenomenological constraints on glueball DM}

In the subsection, we discuss the constraints that exist on glueball DM. They come mainly from the decays of DM and from the strong interactions of glueballs amongst each other.

\subsubsection{Bullet cluster bounds on the scattering rate}

We can estimate the scattering rate among glueballs by 
\bea 
\sigma_{GG \to GG} \sim \frac{220}{\Lambda_{\rm conf}^2}\l(\frac{3}{N}\r)^2 \,. 
\eea 

The bound from the bullet cluster is estimated to be \cite{Randall:2008ppe, Tulin:2017ara}
\bea 
\frac{\sigma_{GG \to GG}}{M_G} \lesssim 2 \times 10^{3} \text{GeV}^{-3} \qquad \Rightarrow \frac{1}{\Lambda_{\rm conf}^3} \lesssim   50 \text{GeV}^{-3} \, ,
\eea 
where we assumed that $M_G \sim 5 \Lambda_{\rm conf}$ \cite{Curtin:2022tou}. 
We thus can conclude that 
\bea 
\Lambda_{\rm conf} \gtrsim 0.25 \text{ GeV} \,. 
\eea 

\subsubsection{Decay of the glueball}

The glueball can decay via different channels depending on its mass. If $\Lambda_{\rm conf}> v$, then the glueballs decay very fast to $h$. So we require that $\Lambda_{\rm conf} < v$. 
 We will consider the constraint on the decay of $G$ to the SM particles, because it is usually more stringent than the decay into some additional dark light particles.

\paragraph{When $M_G < v_{\rm EW}$.}
We first focus on the decay to a SM fermion pair. 
For the reasons advocated in section \ref{production in EWPT}, matching the observed DM abundance requires that $h$ is a BSM Higgs. This BSM Higgs can however always share a quartic portal with the SM, which cannot be forbidden by symmetries. 
Denoting the physical Higgs with a capital letter $H$ and the BSM Higgs with a lower case $h$, we can have
\bea 
\lambda |H|^2 h^2 \; ,
\eea 
 like in \cite{Azatov:2022tii} and the effective operator controlling the decay is then given by 
\beq 
 {\cal L}_{\rm eff}\supset \frac{\lambda v_{\rm EW} v }{m^2_h m^2_H} \frac{v}{\Lambda^2} y_f \bar{f} f G_{\mu \nu}G^{\mu \nu} \, ,
\eeq 
because the mixing term is $\lambda H^2 h^2\supset 2\lambda v_{\rm EW}v H h$. Thus $h$ couples to the  SM fermions via mixing with the Higgs boson $y_e\frac{\lambda v_{\rm EW}v}{m_h^2}$. Taking into account that $m_h \sim v$ and $m_H\sim v_{\rm EW}$, the decay rate becomes
\bea 
\Gamma_{G\to f^+f^-}\sim  \lambda^2\frac{y_f^2 M_G^7}{\Lambda^4 v_{EW}^2} 
\qquad 
t_{G, \rm decay} \sim   \frac{10^{23} s}{y_f^2\lambda^2} \left(\frac{\Lambda}{10^{9}\rm GeV}\right)^4 \left(\frac{0.1\text{ GeV}}{M_G}\right)^7 \, .
\eea 
 The decay is most dominant via the heaviest fermion pair available, i.e. $M_G > 2 m_f = 2 y_f v_{\rm EW}$.   
\paragraph{When $M_G > v_{\rm EW}$.}
In the case of $M_G > 2 m_H$, the glueballs can decay to a pair of Higgses via the operator 
\bea 
 {\cal L}_{\rm eff}\supset \lambda \frac{v^2}{m_h^2} \frac{HH G_{\mu \nu}G^{\mu \nu}}{\Lambda^2}  \, ,
\eea 
which is much less suppressed. 
The decay rate becomes
\bea 
\Gamma_{G\to HH} \sim \lambda^2 \frac{M_G^5}{\Lambda^4} \qquad 
t_{G, \rm decay} \sim \frac{10^{2} s}{\lambda^2} \left(\frac{\Lambda}{10^{9}\rm GeV}\right)^4  \left(\frac{100\text{ GeV}}{M_G}\right)^5 \,. 
\eea 
The bounds on the lifetime of DM impose $t_{G, \text{decay}} \sim 10^{26-27}$s~, e.g.,\cite{Song:2023xdk}.
 Taking the estimate for $\Lambda_{\text{observed DM}}$ in Eq.\eqref{eq:Lambda_est}, we obtain an estimate for the decay time, in the limit of non-supercooling
\bea 
t_{G, \rm decay} \sim \frac{10^{11} s}{\lambda^2\beta^2}   \left(\frac{100\text{ GeV}}{M_G}\right)^{11/3} \bigg(\frac{v}{\text{GeV}} \bigg)^2 \,.
\eea 

\begin{figure}
    \centering
    \includegraphics[width=.41\linewidth]{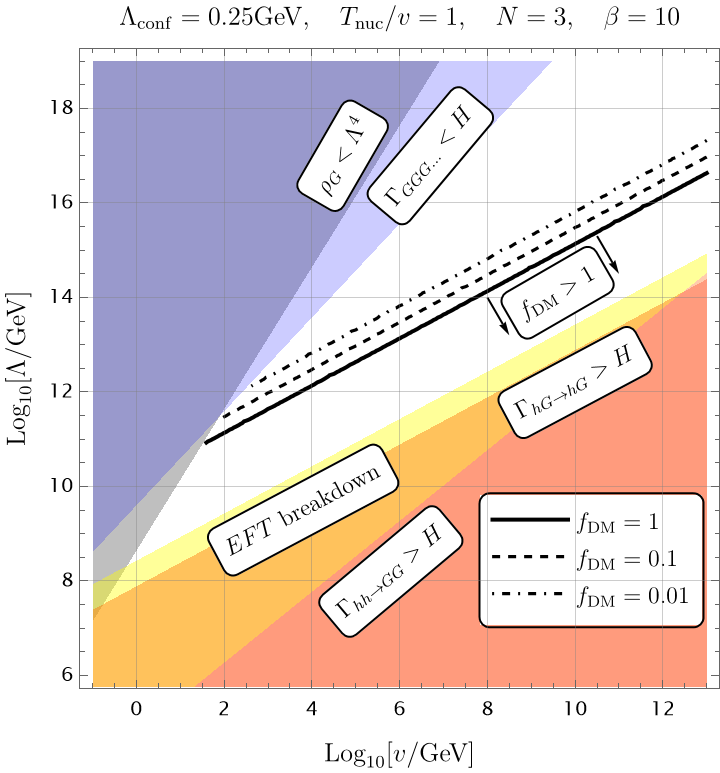} \includegraphics[width=.52\linewidth]{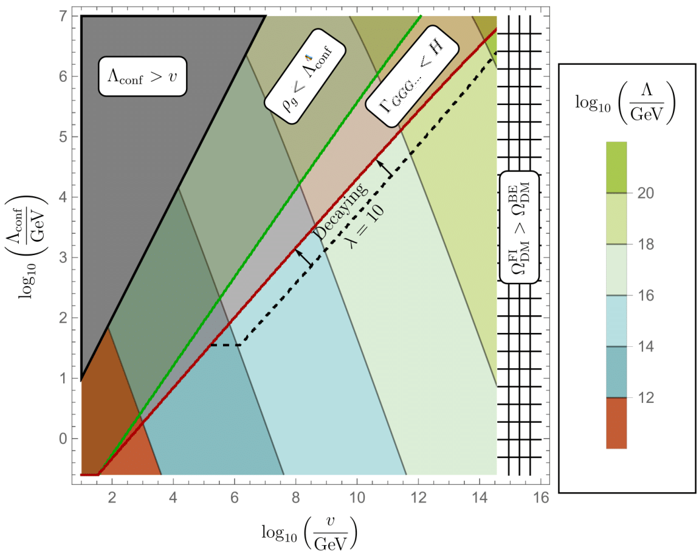}
    
\caption{\textbf{Left}:  We show the different conditions in the specific example $\Lambda_{\rm conf} = 0.25$ GeV. The full black line satisfies the two conditions in $\rho_g > \Lambda_{\rm conf}^4$ and $\Gamma_{GGG...}> H$ and match the observed abundance while the dashed line only underproduce DM for $f_{\rm DM} \equiv \Omega^{\rm BE}_{G}/\Omega_{\rm DM, \text{obs}} = [0.1, 0.01]$. \textbf{Right}: Parameter space allowing to match the observed abundance of DM with the bubble production. 
When $\Gamma_{GGG...}\ll H$, the glueballs are free-streaming, as in the previous discussion for dark vector field. This could contribute to the dark radiation. In the dashed black contour, we present the region of DM that would have decayed by today if $\lambda = 10$.}\label{fig:para_scan_Glueball}
\end{figure}

In Fig. \ref{fig:para_scan_Glueball}, we present the region of parameter space allowing to match the observed abundance of DM from bubble expansion assuming the melted regime.
On the Left panel, the black (full, dashed and dash-dotted) lines show the region matching the observed abundance of DM (or a fraction of it in dashed lines), in the melted regime. We shade all the regions where our computation is not valid: the yellow region excludes the region where the EFT is broken, according to Eq.\eqref{eq:EFT_cond}, the orange region shows the region where the $h G \to h G$ interaction is \emph{active}, according to Eq.\eqref{eq:rate_phi_g}, and the red region is the region where $hh \to G G$ is active. Those reactions would couple the thermal sector to the glueball sector and induce an unacceptably large abundance of glueballs. 
Finally, the blue region shows the region where the interaction $GG \to GG...$ is \emph{inactive} according to Eq.\eqref{eq:GG_active} with the rates for FI and BE given by Eq.\eqref{eq:GG_from_FI} and \eqref{eq:GG_from_BE}, 
respectively, and the gray region denotes the region where the requirements of having enough energy to form a gluon plasma is not fulfilled, i.e. Eq.\eqref{eq:qg_enough_E}. In the latter two regions, we do not expect the dark sector to ever reach a deconfined gluon phase and then our computation is likely not valid. 

On the Right panel of Fig.\ref{fig:para_scan_Glueball}, we observe that successful glueball DM can occur in a vast region of the parameter space 
\bea 
v \in [10^4, 10^{15}] \text{ GeV}, \qquad \Lambda \in [10^{13}, 10^{19}] \text{ GeV}, \qquad \Lambda_{\rm conf} \in [0.25, 5 \times 10^6] \text{ GeV} \,. 
\eea 
For almost all the range of the strongly coupled sector that matches the DM abundance, the DM is stable on cosmological timescales.  However in a thin band of the parameter space, for large glueball masses, and for a large portal coupling $\lambda \approx 10$, the DM is decaying. 

At the other edge of the parameter space, for light glueballs, the strong interactions among DM glueballs can show up in bullet cluster-like events if $\Lambda_{\rm conf} \sim 0.3$ GeV, hence we use this value as a lower limit for the range of the parameter scan.\\

\section{GW signal from Dark Matter production}
\label{sec:GW_signal}
In the previous sections we studied the production of DM during the phase transition. We saw that it required runaway walls and rather slow transitions with moderate values of $\beta$. In turn this specific type of phase transition is expected to induce a large background of gravitational waves due to the sound waves and the collision of bubble walls, making this mechanism possibly detectable via GW interferometers. 

\begin{figure}
    \centering
    \includegraphics[width=.49\linewidth]{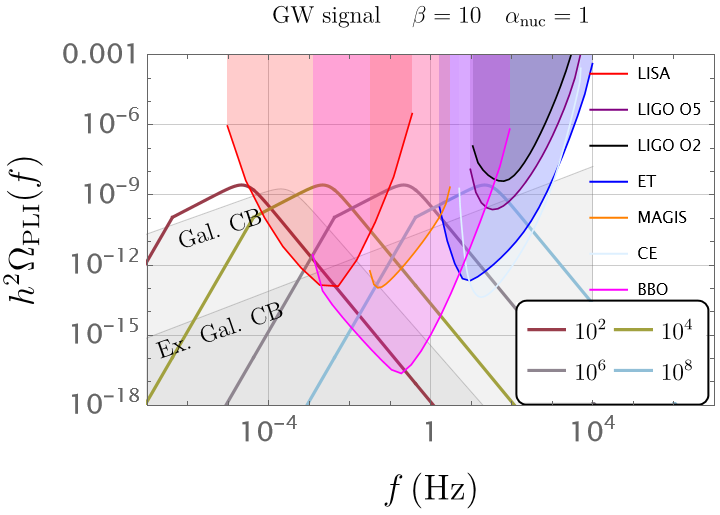}
    \includegraphics[width=.49\linewidth]{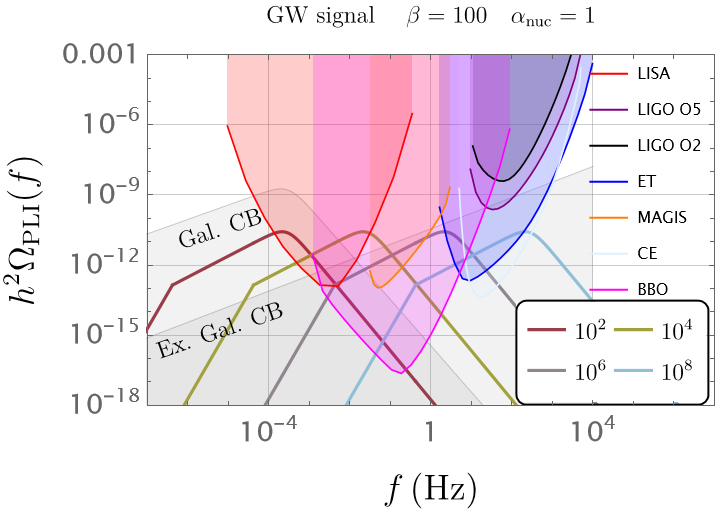}
    \caption{Shape of the GW signal for the model producing DM with $\alpha_{\rm nuc} =1$ and $\beta = 10$ (Left) and $\beta= 100$ (Right) for different values of $v = 10^2, 10^4, 10^6, 10^8$ GeV. The signal-to-noise ratio and the sensitivity curves can be build following the recommendations of\cite{Moore:2014lga,Aasi:2013wya,TheLIGOScientific:2014jea,Cornish:2018dyw,Graham:2017pmn,Yagi:2011yu,Yagi:2013du,Sathyaprakash:2012jk}. The gray regions are the expected foregrounds from galactic and extra-galactic compact binaries that we obtain from the recommendations in \cite{KAGRA:2021kbb,Boileau:2021gbr,Boileau:2022ter}.  }
    \label{fig:GW_Bulk}
\end{figure}

In this section, we quickly comment on this GW background. In the simplest version of our model, the energy of the phase transition goes to the bubble wall stress, the shells of produced DM and strongly boosted fluid shells.   For such a fast bubble expansion  the bulk flow model \cite{Konstandin:2017sat}
\footnote{The authors thank Jorinde Van De Vis and Ryusuke Jinno for helpful discussions on the bulk flow model.} is expected to describe the stochastic GW background the best \cite{Baldes:2024wuz} and we use it to assess the current and future experimental  sensitivities.
We present the signal of the GW signal induced together with the integrated power-law sensitivities of LISA, LIGO, CE, ET, MAGIS, BBO in Fig.\ref{fig:GW_Bulk}.
Following Ref. \cite{Konstandin:2017sat}  the GW signal, assuming $v_w \to 1$, takes the form
\bea 
h^2\Omega_{GW} = h^2\Omega_{\rm peak} S(f, f_{\rm peak}) \qquad 
S(f, f_{\rm peak}) = \frac{(a+b) f_{\rm peak}^b f^a}{b f_{\rm peak}^{(a+b)}+ a f^{(a+b)} }, \qquad (a, b) \approx (0.9, 2.1) \, , 
\eea 
with
\begin{align}
h^2\Omega_{\rm peak} &\approx 1.06 \times 10^{-6} \bigg(\frac{H}{\beta}\bigg)^2 \bigg(\frac{\alpha_{\rm nuc}\kappa}{1+\alpha_{\rm nuc}}\bigg)^2 \bigg(\frac{100}{g_\star}\bigg)^{1/3} \qquad \text{and} \qquad  \kappa =1 \, ,
\nn
f_{\rm peak} &\approx 2.12 \times 10^{-3} \bigg(\frac{\beta}{H_{\rm reh}}\bigg)\bigg(\frac{T_{\rm reh}}{100 \text{GeV}}\bigg) \bigg(\frac{100}{g_\star}\bigg)^{-1/6} \quad \text{mHz} \, .
\end{align}
Note that on the top of this spectrum, one needs to impose an IR cut-off required by causality for $f < H_{\rm reh}/2\pi$.

The complicated problem of the separation of the unavoidable astrophysical background (from time to time called \emph{foreground}) from the 
possible cosmological background is still under vivid investigation \cite{Caprini:2019pxz,Flauger:2020qyi,Boileau:2020rpg, Martinovic:2020hru}. As a 
consequence, we shade in gray the regions where we expect a strong foreground from galactic and extra-galactic compact binaries. This foreground is still subject to very large uncertainties and will depend on our abilities to resolve individual sources, it should therefore be interpreted with caution.

\section{Summary and conclusions}
\label{sec:summary}
In this paper, we have studied the bubble wall production of DM in a secluded sector connected to the thermal bath by a non-renormalisable portal. We have first systematically studied the abundance of DM, the spectrum of the emitted DM and its evolution for a dimension five (fermion) and a dimension six portal (vector) weakly coupled DM scenarios. 
Unfortunately, both scenarios are incompatible with EW phase transition, since the bubble expansion will not be sufficiently fast due to the gauge boson friction.
Requiring that the BE production matches the observable abundance, we observe that very interestingly a phase transition at scale close to the EW scale, $v \sim 10^2 - 10^3$ GeV, can lead to DM masses up to $10^{10}$ GeV. We observed that in both cases, of vector and fermion production, the EFT validity constraint requires that the DM is generically \emph{free-streaming} after 
the production. On the other hand, both the dimension five and the dimension six operators allow in a band of the parameter space the production of WDM with $V_{\rm eq} \sim 10^{-5}$ or slightly faster.  
We found that in the instance of an electroweak scale PT, on  top of the expected GW signal accompanying the strong PT, the 
bubble wall production of secluded warm fermionic and vector DM will be WDM 
and have of large mass of order $M_{(\psi, \gamma)} \sim 10^8- 10^9$ GeV. This is for example 
illustrated by the red star on Fig.\ref{fig:dim5_money_1}. Those signatures constitute a striking smocking gun of our scenario.

In parallel, we have calculated the pressure on the bubble wall caused 
by  higher dimensional operators.
Reassuringly, this new contribution to the pressure remains minor across all the parameter spaces we examined, making it unlikely to significantly alter the velocity of the wall.

We also studied the case of a pure Yang-Mills confining dark sector, where the DM is composed of dark glueballs. In this scenario, we focused on the production of a thermal dark 
gluon  plasma. Here, the energy of each individual boosted glueball,  initially  produced by the wall, contributes  
to increasing the final number density of glueballs through the self-interactions of the dark sector.
Consequently, glueball DM is always slow.
The formation of a gluon plasma, combined with the Bullet Cluster constraint on DM self-interactions, requires the mass of the glueball to be within the range of $[0.25, 5\times 10^6]$ GeV.

Finally, we emphasize that our proposal for DM production is inherently 
accompanied by strong signals in the stochastic GW background. This makes 
our proposal highly detectable.

\section*{Acknowledgements}

It is a pleasure to thank Iason Baldes for comments on the manuscript. 
XN is supported by the iBOF “Un-locking the Dark Universe with Gravitational Wave Observations: from Quantum Optics to Quantum Gravity” of the Vlaamse Interuniversitaire Raad and the "Strategic Research Program High-Energy Physics of the Vrije Universiteit Brussel”. MV is supported by the ``Excellence of Science - EOS" - be.h project n.30820817, and by the Strategic Research Program High-Energy Physics of the Vrije Universiteit Brussel. 
 WY is supported by JSPS KAKENHI Grant Nos.  20H05851, 21K20364, 22K14029,  22H01215, and 23K22486.
AA is supported in part by the MUR projects 2017L5W2PT. AA also acknowledges support by the European Union - NextGenerationEU, in the framework of the PRIN Project “Charting unexplored avenues in Dark Matter" (20224JR28W). 

\appendix

\section{Bubble-Plasma production probability}
\label{App:prodDM}

In this appendix, we gather the details of the computation of the production via Bubble-Plasma interactions that we have studied in the main text. 

\subsection{Dimension five operator}
In the main text, we studied the impact of the following dimension 5 operator 
\bea 
\frac{h^2 \psi \bar \psi}{\Lambda} \, ,
\eea 
where $\Lambda$ is the UV cutoff, $h$ undergoes the phase transition $ h \to h+v$ and $\psi$ is a fermion of mass $M_\psi$. 
We present the details of the computation of the probability of $h \to \psi \psi$ production. 
The kinematics of the process can be parameterized as follows
\bea
\label{Eq:kinematics}
p_a&=&(p_0, 0, 0, p_0),
\nn 
p_b&=&((1-x) p_0, k_\perp, 0, \sqrt{(1-x)^2p_0^2-M_\psi^2- k_\perp^2}),
\nn
p_c&=&(x p_0, -k_\perp, 0, \sqrt{x^2p_0^2-M_\psi^2- k_\perp^2}) \, , 
\eea 
where $M_\psi < \Lambda$ is the mass of the $\psi$ which is constant.
From this point we can thus introduce a WKB wave for the outgoing particles X, as presented in \cite{Bodeker:2017cim} and reviewed in \cite{Azatov:2020ufh}. Our computation is typically valid in the wall frame for very large velocities: $p_z \sim \gamma_w T_{\rm nuc} \gg 1/L_w$. Within this WKB approach, the matrix element takes the form
\bea 
\label{Eq:vertex}
\mathcal{M} = i\bigg(\frac{V_s}{A_s}- \frac{V_h}{A_h}\bigg)  \, ,
\eea 
where the $s(h)$ subscript denotes the symmetric(higgsed) side. So 
\bea 
A \equiv p_0 - \sqrt{(1-x)^2p_0^2-M_\psi^2- k_\perp^2} - \sqrt{x^2p_0^2-M_\psi^2- k_\perp^2} \,. 
\eea 
and the squared vertex $|V^s_{h \to \psi \psi}|^2 =0$ and\footnote{we can understand this intuitively by the fact that the vertex $\propto v h \psi \bar \psi$ only exists on the higgsed, broken side.}
\begin{align}
|V^h_{h \to \psi \psi}|^2 &= 4 \frac{v^2}{\Lambda^2} (p_b \cdot p_c -M_\psi^2) 
\nn
&= 4 \frac{v^2}{\Lambda^2} \bigg(p_0^2 x(1-x) - p_0^2x (1-x) \sqrt{1- \frac{M_\psi^2+k_\perp^2}{p_0^2x^2}}\sqrt{1-\frac{M_\psi^2+k_\perp^2}{p_0^2(1-x)^2}} + k_\perp^2 - M_\psi^2\bigg)
\nn
&=2 \frac{v^2}{\Lambda^2}\frac{1}{x(1-x)}\left(M_\psi^2(2x-1)^2+k_\perp^2\right) \;.
\end{align}
Expanding for $x^2(1-x)^2p_0^2 \gg k_\perp^2 + M_\psi^2$, we obtain
\bea 
A_h = A_s \approx \frac{M_\psi^2+k_\perp^2}{2 p_0(1-x)x}  \,. 
\eea

The amplitude matrix then becomes 

\bea
\label{Eq:matrixEl}
 |\mathcal{M}_{h \to \psi \bar \psi}|^2=8\frac{v^2}{\Lambda^2}\left(\frac{p_0}{M_\psi}\right)^2\frac{x(1-x)}{1+\left(\frac{k_\perp}{M_\psi}\right)^2}\left(1-\frac{4x(1-x)}{1+\left(\frac{k_\perp}{M_\psi}\right)^2}\right) \, .
\eea 
 
The value of $k_\perp^{\rm max}$ can be straightforwardly extracted from the non-adiabaticity condition. For more details on how to recover this conditions from basic principles, see Appendix A of \cite{Azatov:2021ifm} and Appendix H of \cite{Azatov:2024auq} for further discussions. Therefore, the upper boundary of the $k_\perp$ integral reads:
 \bea 
 \label{Eq:non-adia}
\frac{M_\psi^2+k^2_\perp}{2v p_0}< (1-x)x \qquad \text{(non-adiabaticity condition)} \,.
\eea 
which implies that the upper bound on the integral is given by 
\bea 
\label{Eq:adiab}
(k_\perp^{\rm max})^2=2vp_0x (1-x) - M_\psi^2  \approx 2vp_0x (1-x)\, .
\eea
 Using the matrix element in Eq.\eqref{Eq:matrixEl} and the boundaries in Eq.\eqref{Eq:adiab} and integrating over the phase space, the probability of producing $h \to \psi \psi$ is
\begin{align}
    P_{h \to \psi \psi} &\approx \frac{1}{8p_0^2}   \int_{b_+}^{b_-} \frac{dx}{x(1-x)} \int^{ 2 p_0 v x (1-x) -M_\psi^2}_0\frac{dk_\perp^2}{8\pi^2}  |\mathcal{M}_{h \to \psi \bar \psi}|^2 
    \nn 
     &\approx   \frac{v^2}{8\pi^2\Lambda^2}\left[\frac{4}{3}\sqrt{1-\frac{2M_\psi^2}{p_0v}}\left(\frac{M_\psi^2}{2p_0v}-1\right)+\log\left(\left|1-\frac{p_0 v}{M_\psi^2}-\sqrt{\left(\frac{p_0 v}{M_\psi^2}-2\right)\frac{p_0 v}{M_\psi^2}}\right|\right)\right] \,.
\end{align}
and we defined
\bea 
b_+ \equiv \frac{1-\sqrt{1-\frac{2M_\psi^2}{p_0v}}}{2}, \quad b_- \equiv \frac{1+\sqrt{1-\frac{2M_\psi^2}{p_0v}}}{2} \, .
\eea

\subsection{Dimension six operator}

Let us now consider the dimension 6 operator between scalars and gauge bosons of the form
\bea 
\frac{h^2 F_{\mu \nu}F^{\mu \nu}}{\Lambda^2} \, ,
\eea 
where $F$ is the field strength of a dark vector $\gamma$. In this section we study the splitting $h \to \gamma \gamma$, the kinematics are given as usual by 
\bea
p_a&=&(p_0, 0, 0, p_0\,),\\
p_b&=&((1-x) p_0, k_\perp, 0, \sqrt{(1-x)^2p_0^2- k_\perp^2-M_\g^2}),\\
p_c&=&(x p_0, -k_\perp, 0, \sqrt{x^2p_0^2- k_\perp^2-M_\g^2}),
\eea 

We will work in the unitary gauge where the ghosts and the Goldstone boson have an infinite mass and are thus decoupled from the theory. We are thus left with only massive gauge bosons, two transverse modes $\pm$ and one longitudinal mode $0$. 
We first consider the two transverse  polarization vectors, which are given by 
\begin{align}
    \epsilon^\pm_b&=\frac{1}{\sqrt{2-2\frac{M_\g^2}{(1-x)^2p_0^2}}}\left(0,\sqrt{1-\frac{k_\perp^2+M_\g^2}{(1-x)^2p_0^2}},\pm i\sqrt{1-\frac{M_\g^2}{(1-x)^2p_0^2}},-\frac{k_\perp}{(1-x)p_0}\right) \, ,
\nn
    \epsilon^\pm_c &=\frac{1}{\sqrt{2-2\frac{M_\g^2}{x^2p_0^2}}}\left(0,\sqrt{1-\frac{k_\perp^2+M_\g^2}{x^2p_0^2}},\pm i\sqrt{1-\frac{M_\g^2}{x^2p_0^2}},\frac{k_\perp}{x p_0}\right) \, .
\end{align}
 Since in the frame transition $p_0 \gg M_\psi$, we can apply a massless limit and one obtains the following scalar products:
\begin{align}
\label{eq:pol_sum}
 p_b.\epsilon_b &=0 \, ,
\\
    \epsilon_b^-\cdot \epsilon_c^-&\approx\epsilon_b^+\cdot \epsilon_c^+ \approx \frac{k^2_\perp}{4 p^2_0x^2(1-x)^2} \to 0 \, ,
    \\
    \epsilon_b^-\cdot \epsilon_c^+&\approx \epsilon_b^+\cdot \epsilon_c^- \approx  -1 + \frac{k^2_\perp}{4 p^2_0x^2(1-x)^2} \to -1 \, ,
    \\
    p_{b}\cdot\epsilon^{+/-}_c &\approx -\frac{k_\perp}{\sqrt{2}x}+\frac{k_\perp(k_\perp^2 x-M_\g^2 (1-2x))}{2\sqrt{2}x^3(1-x)p_0^2} \to -\frac{k_\perp}{\sqrt{2}x} \, ,
    \\
    p_{c}\cdot\epsilon^{+/-}_b &\approx \frac{k_\perp}{\sqrt{2}(1-x)}-\frac{k_\perp(k_\perp^2(1-x)+M_\g^2 (1-2x))}{2\sqrt{2}x(1-x)^3p_0^2} \to \frac{k_\perp}{\sqrt{2}(1-x)} \, ,\\
    p_b\cdot p_c &\approx \frac{k_\perp^2+M_\g^2 (1-2x(1-x))}{2x(1-x)}+\frac{(k_\perp^2+M_\g^2)^2(1-2x)^2}{8(1-x)^3x^3p_0^2} \to \frac{k_\perp^2+M_\g^2 (1-2x(1-x))}{2x(1-x)} \, ,
\end{align}
which simplifies in the limit $p_0 \gg k_\perp > M_\psi$ by the quantity designated by the arrow. 
As before the vertex on the symmetric side is zero while the vertex on the higgsed side takes the form
\bea 
V^{\lambda \lambda'}_h = \frac{2v}{\Lambda^2} ((p_b \cdot p_c) (\epsilon_b^\lambda \cdot \epsilon_c^{\lambda'}) - (p_c\cdot \epsilon_b^{\lambda'})(p_b\cdot \epsilon_c^\lambda))  \, . 
\eea 
which gives after plugging Eq.\eqref{eq:pol_sum}
\bea 
V^{++}_h = V^{--}_h = \frac{v}{\Lambda^2}\frac{k_\perp^2}{x (1-x)}, \qquad V^{+-}_h = V^{+-}_h = -\frac{v}{\Lambda^2}\frac{M_\g^2 (1-2x(1-x))}{x (1-x)} \, . 
\eea 

Keeping only the leading order terms in $k_\perp^2/p_0^2$, we obtain the following expression for the vertex function squared and summing over the polarisations, we obtain
\begin{align}
    |V_h|^2 \equiv |V^{++}_h|^2 + |V^{--}_h|^2 +2|V^{+-}_h|^2\approx\frac{v^2}{\Lambda^4}\frac{2}{x^2(1-x)^2}\left(k_\perp^4+M_\g^4(1-2x(1-x))^2\right) \quad ,
\end{align}
and consequently, using Eq.\eqref{Eq:vertex}, we have 
\begin{align}
\label{Eq:matrix_el_dim_6}
    |\mathcal{M}_{h \to \gamma \gamma}|^2 \approx \frac{8(p_0 v)^2}{(M_\g^2+k_\perp^2)^2\Lambda^4}\left(k_\perp^4+M_\g^4(1-2(1-x)x)^2\right) \approx \frac{8(p_0 v)^2k_\perp^4}{(M_\g^2+k_\perp^2)^2\Lambda^4} \, ,
\end{align}
for negligible $M_\gamma$. As before, the probability of emission of tranverse gauge bosons is obtained in the following way 
\begin{align}
P^T_{h \to \gamma \gamma}
&\approx    \frac{1}{8p_0^2}   \int_{b_+}^{b_-} \frac{dx}{x(1-x)} \int^{ 2 p_0 v x (1-x)-M^2_\g}_0\frac{dk_\perp^2}{8\pi^2}  |\mathcal{M}|^2\\
&\approx \frac{v^2M_\g^2}{8\pi^2\Lambda^4}\int_{b_+}^{b_-} \frac{dx}{x(1-x)} \int^{ 2 p_0 v x (1-x)/M^2_\g}_0d\left(\frac{k_\perp^2}{M_\g^2} \right) \frac{(k_\perp^4/M_\g)^4}{(1+(k_\perp/M_\g)^2)^2} \quad .
\end{align}
where we defined 
\bea 
b_+ \equiv \frac{1-\sqrt{1-\frac{2M_\g^2}{p_0v}}}{2}, \quad b_- \equiv \frac{1+\sqrt{1-\frac{2M_\g^2}{p_0v}}}{2} \, .
\eea 
The function in the last integrand tends rapidly to one and we can therefore approximate it with unity so that the expression becomes
\begin{align}
  P_{h \to \gamma \gamma}  \approx \frac{2 p_0v^3}{8\pi^2\Lambda^4}\int_{b_+}^{b_-} dx  \approx \frac{v^3p_0}{4\pi^2\Lambda^4}\sqrt{1-\frac{2M_\g^2}{p_0v}}\to \frac{v^3p_0}{4\pi^2\Lambda^4} \quad , 
\end{align}

    Incidentally, the production of gluons from $h$ via 
\bea 
\frac{h^2 G_{\mu \nu}G^{\mu \nu}}{\Lambda^2}
\eea 
is of the same form 
\bea 
P_{h \to gg}= P_{h \to \gamma \gamma} \quad .
\eea

\section{Pressure on the bubble wall from particles production}
\label{app_pressures}

In Ref. \cite{Azatov:2020ufh}, it was pointed out that the production of heavy particles by the bubble wall is accompanied by an exchange of momentum from the plasma to the bubble: each produced pair of $\psi$ particles act as a kick on the bubble wall, which appears as an effective pressure on the bubble wall when we integrate over the incoming flux. 
In this appendix, we recompute the pressure in the relativistic regime induced by particles produced via $h \to XX$. 

    Lorentz violating interactions will transfer momentum from the plasma to the wall. This \emph{momentum} exchange can be identified with the loss of the momentum from the emitted particles, which in a $1 \to 2$ process, takes the form 
    \bea 
    \Delta p_z = p_a^z- p_b^z - p_c^z \, ,
    \eea 
and can be evaluated from the kinematics in Eq.\eqref{Eq:kinematics}. The pressure on the wall can thus be identified with the convolution of this exchange of momentum and the probability of the interaction, together with the incoming flux,
\bea 
\label{eq:pressure_generic}
\mathcal{P}_{a \to bc} = \underbrace{\int \frac{d^3p}{(2\pi)^3} f_h (p_a)}_{\text{incoming flux}}  \int dP_{a \to bc} \times \Delta p_z . 
\eea 
Using now the probability of the emission $h \to XX$,
\bea 
dP_{a \to bc} = \frac{d^3 p_b d^3 p_c}{(2\pi)^6 2 E_b 2 E_c} |\mathcal{M}_{a \to bc}|^2 (2\pi)^3 \delta^2 \bigg(\sum_i p^i_\perp \bigg)\delta\bigg(\sum_iE^i \bigg) \, , 
\eea 
we will compute this integral for the different cases we have considered in the main text.

\subsection{Dimension five operator}
Let us first consider the computation of the pressure from the production of $X = \psi$ dark matter. In the regime of a fast bubble wall $(1-x)^2x^2 p_0^2 \gg M_\psi^2 + k_\perp^2$, 
\bea 
\Delta p_z \approx \frac{M_\psi^2 + k_\perp^2}{2 x (1-x) p_0} \, ,
\eea 
and the pressure can be computed with the following steps 
\begin{align}
\mathcal{P}_{h \to \psi \psi}
&\approx  2 \frac{v^2}{\Lambda^2} \int \frac{d^3 p}{(2\pi)^3} f_h(p) \int^{2 p_0 v x (1-x)}_0 \frac{dk^2_\perp}{8\pi^2}  \int^{1-k_\perp/p_0}_{k_\perp/p_0} \frac{dx}{(M_\psi^2+k_\perp^2)} \times \frac{M_\psi^2 + k_\perp^2}{2 x (1-x) p_0}
\nn
&= \frac{v^3}{\Lambda^2} \int \frac{d^3 p}{(2\pi)^3} f_h(p) \frac{1}{8\pi^2} \, ,
\end{align}
which finally leads to 
\bea 
\mathcal{P}_{h \to \psi \psi}\approx  \frac{1}{8\pi^2}\frac{v^3 n_h \gamma_w}{\Lambda^2}  \,. 
\eea

\subsection{Dimension six operator}

We now turn to the pressure induced by the dimension six operator. We can easily estimate the pressure exerted on the bubble by multiplying with the production probability from Eq.\eqref{Eq:matrix_el_dim_6} with exchange of momentum $\Delta p_z$ and using the results in Eq.\eqref{eq:pressure_generic}, we obtain
\begin{align}
\mathcal{P}_{h \to \gamma \gamma } &\approx \int \frac{d^3p}{(2\pi)^3} f_h(p) \frac{1}{4p_0^2}\int^{2 v p_0x (1-x)}_0\frac{dk_\perp^2}{8\pi^2}   \int_{k_\perp/p_0}^{1- k_\perp/p_0} \frac{dx}{x(1-x)}  \frac{k_\perp^2+ M_\gamma^2}{2x(1-x)p_0} \times \frac{8(p_0 v)^2k_\perp^4}{(M_\g^2+k_\perp^2)^2\Lambda^4} 
\nn 
 &\approx 
\int \frac{d^3p}{(2\pi)^3} \frac{v^4}{\Lambda^4}   \frac{p_0}{2\pi^2} f_h(p)
 \nn
 &\approx n_h \frac{v^4}{\Lambda^4}   \frac{\gamma_w^2 T}{2\pi^2}   \,.  
\end{align} 

As for the particle production, the pressure is identical for vectors and for gluons 
\bea 
\mathcal{P}_{h \to gg} = \mathcal{P}_{h \to \gamma \gamma} \,. 
\eea 

\section{Computation of the Freeze-In abundance of DM}
\label{App:FI_abundances}
An irreducible contribution to the produced DM abundance comes from the high temperature Freeze-In via $2 \to 2$ scatterings. In this Appendix, we provide our estimates for the FI abundance via the different production channels that we considered in the main text. 

\subsection{Dimension five operator}
If the $h$ particles are in equilibrium with the thermal bath, the light DM particle will undergo Freeze-In via the $2\to 2$ scatterings $hh \to \psi \bar \psi$. The scattering matrix is given by 
$|\mathcal{M}_{hh \to \psi \psi}|^2 = \frac{2}{\Lambda^2}(s- 4 M_\psi^2)$
and the Boltzmann equation for this production mechanism takes the form \cite{Hall:2009bx}:
\begin{align}
\dot n_\psi + 3 H n_\psi = a^{-3} \frac{d (a^3 n_\psi)}{dt} &= \frac{T}{512 \pi^6} \int_{s_{\rm min}}^{\infty} d\Omega ds \frac{P_{hh}P_{\psi \psi}}{\sqrt{s}} |\mathcal{M}_{hh \to \psi \psi}|^2 K_1 (\sqrt{s}/T)
\\ \nonumber
&\approx  \frac{T^5M_\psi}{8 \pi^5  \Lambda^2} \times    \sqrt{\frac{\pi T}{M_\psi}} e^{-2M_\psi/T}
\end{align}
where we defined 
$P_{ij} \equiv \frac{\sqrt{s-(m_i+m_j)^2}\sqrt{s-(m_i-m_j)^2}}{2 \sqrt{s}}$, with $P_{hh} \approx \sqrt{s}/2$ and $P_{\psi \psi} \approx \sqrt{s-4 M_\psi^2}/2$. 
To obtain an analytical answer we applied the following approximation
$ (s-4M^2)^{3/2} =  (s-4M^2)  (s-4M^2)^{1/2} \approx  (s-4M^2)4T$ and we have checked that this is in good agreement numerically. In the second line we took the limit of $M_\psi \gg T$, which we will follow from now on. 
We have also approximated the $zK_1(z) \to \sqrt{\pi z/2} e^{-z} \approx \sqrt{\pi M_\psi /T} e^{-z}$. The density normalized to the entropy density, $Y_\psi^{\rm FI}= n_\psi^{\rm FI}/s$ will hence become
\begin{align}
    & a^{-3}\frac{d(a^3n_\psi)}{dt}=-H T^4 \frac{d \left( \frac{n_\psi}{T^3}\right)}{dT}=-HT^4 g_\star \frac{2\pi^2}{45} \frac{dY}{dT}
    \approx  \frac{T^5M_\psi}{8 \pi^5  \Lambda^2} \times    \sqrt{\frac{\pi T}{M_\psi}} e^{-2M_\psi/T}\nonumber\\
    \Leftrightarrow& \qquad \frac{d Y}{dT}\approx -\frac{45 M_{\rm pl}}{16 g_\star^{3/2} \pi^7 1.66 \Lambda^2} \times    \sqrt{\frac{\pi M_\psi}{T}} e^{-2M_\psi/T}\nonumber \\
    \Leftrightarrow& \qquad  Y_\psi^{\rm FI} \approx \frac{45 M_{\rm pl}}{16 g_\star^{3/2} \pi^{13/2} 1.66 \Lambda^2} \times \frac{M_\psi}{2}    \left(\frac{ T_{\rm reh}}{ M_\psi}\right)^{3/2}e^{-2M_\psi/T_{\rm reh}} \quad ,
\end{align}
where the last line is a fit of our numerical analysis. 
 This translates into a DM fraction today of 
\begin{align}
\Omega^{\rm FI}_{\psi, \rm today} h^2  &= \frac{M_\psi Y^{FI}_{\psi} s_0}{\rho_c/h^2} 
\nonumber \\ \nonumber
&= 2.35 \times 10^8 \bigg(\frac{M_\psi}{\text{GeV}} \bigg) \frac{45 M_{\rm pl}}{16 g_\star^{3/2} \pi^{13/2} 1.66 \Lambda^2} \times \frac{M_\psi}{2}    \left(\frac{ T_{\rm reh}}{M_\psi}\right)^{3/2}e^{-2M_\psi/T_{\rm reh}}
\\ 
&= 5.84 \times 10^4 \bigg(\frac{M_\psi}{\text{GeV}} \bigg) \frac{ M_{\rm pl} M_\psi }{ g_\star^{3/2} \Lambda^2  } \left(\frac{ T_{\rm reh}}{M_\psi}\right)^{3/2} e^{-2M_\psi/T_{\rm reh}} \,. 
 \end{align}

\subsection{Dimension six operator: vectors}
Let us repeat the same computation for the FI of massive vectors, the computation will follow the same steps.
The scattering matrix for the vector portal is now given by 
\bea 
|\mathcal{M}_{hh \to \gamma \gamma}|^2 \approx  \frac{1}{\Lambda^4}\left(2(s- 2M_\g^2)^2+4M_\g^4\right) \,. 
\eea 
 We will perform the computation in the two following limits: $M_\gamma \gg T$ and $M_\gamma \ll T$. 

\paragraph{In the case,  $M_\gamma \gg T$}:
The matrix element takes the form 
\bea 
|\mathcal{M}_{hh \to \gamma \gamma}|^2 \approx  \frac{1}{\Lambda^4}\left(2(s- 2M_\g^2)^2+4M_\g^4\right) \approx \frac{12M_\gamma^4}{\Lambda^4}
\eea 
where the second approximation comes from the fact that we consider the regime $M_\gamma \gg T$ where the exponential suppression selects the smallest value of the $s \approx  4M_\gamma^2$.
The Boltzmann equations for this production take the form 
\begin{align}
\dot n_\gamma + 3 H n_\gamma = a^{-3} \frac{d (a^3 n_\gamma)}{dt} &= \frac{T}{512 \pi^6} \int_{s_{\rm min}}^{\infty} d\Omega ds \frac{P_{hh}P_{\gamma \gamma}}{\sqrt{s}} |\mathcal{M}_{hh \to \gamma \gamma}|^2 K_1 (\sqrt{s}/T)
\\ \nonumber
&\approx  \frac{3 T^4 M^4_\g}{32\pi^{9/2}   \Lambda^4} \times \sqrt{\frac{M_\gamma}{T}} e^{-2M_\gamma/T} \,, 
\end{align}

where we again applied 
$ (s-4M_\g^2)^{1/2} \approx 2T$.
It follows that 
\begin{align}
    \frac{dY_\g^{\rm FI}}{dT}&\approx \frac{135 M_{\rm pl} M^4_\g}{64\pi^{13/2}g_\star^{3/2} 1.66 T^2\Lambda^4} \sqrt{\frac{M_\gamma}{T}} e^{-2M_\gamma/T}
    \nonumber \\
    Y_\g^{\rm FI} &\approx \frac{135 M_{\rm pl} M^3_\g}{128\pi^{13/2}g_\star^{3/2} 1.66 \Lambda^4} \sqrt{\frac{M_\gamma}{T_{\rm reh}}} e^{-2M_\gamma/T_{\rm reh}} \,.
\end{align}

Finally 
\begin{align}
\Omega^{\rm FI}_{\gamma, \rm today} h^2  &= \frac{M_\gamma Y^{FI}_{\gamma} s_0}{\rho_c/h^2} \nonumber
\\ 
&\approx 2.35 \times 10^8 \bigg(\frac{M_\gamma}{\text{GeV}} \bigg)\frac{135 M_{\rm pl} M^3_\g}{128\pi^{13/2}g_\star^{3/2} 1.66 \Lambda^4} \sqrt{\frac{M_\gamma}{T_{\rm reh}}} e^{-2M_\gamma/T_{\rm reh}} \nonumber \\
&\approx 8.76 \times 10^4 \bigg(\frac{M_\gamma}{\text{GeV}} \bigg)\frac{M_{\rm pl} M^3_\g}{g_\star^{3/2} \Lambda^4} \sqrt{\frac{M_\gamma}{T_{\rm reh}}} e^{-2M_\gamma/T_{\rm reh}} \,. 
 \end{align}

\paragraph{In the case,  $M_\gamma \ll T$}: In the case of relativistic FI, we obtain

\bea 
|\mathcal{M}|_{hh \to \gamma \gamma}^2
\simeq \frac{2 s^2 }{\Lambda^4} \qquad a^{-3} \frac{d (a^3 n_\gamma)}{dt} \approx   \frac{3T^8 }{\pi^5 \Lambda^4} \, , 
\eea 
which leads to 
\bea 
\label{eq:FI_rel}
Y^{M_\gamma \ll T}_{\rm FI} \approx  \frac{45 M_{\rm pl} T_{\rm reh}^3}{3.32 \pi^7 g_*^{3/2} \Lambda^4   }  \, ,
\eea 
and 
Finally 
\begin{align}
Y^{M_\gamma \ll T}_{\rm FI}  h^2  &= \frac{M_\gamma Y^{FI}_{\gamma} s_0}{\rho_c/h^2} \nonumber
\\ 
&\approx 2.35 \times 10^8 \bigg(\frac{M_\gamma}{\text{GeV}} \bigg)\frac{45 M_{\rm pl} T_{\rm reh}^3}{3.32 \pi^7 g_*^{3/2} \Lambda^4   } \nonumber \\
&\approx 1.05 \times 10^6 \bigg(\frac{M_\gamma}{\text{GeV}} \bigg)\frac{M_{\rm pl} T_{\rm reh}^3}{g_\star^{3/2} \Lambda^4} \,. 
 \end{align}

\subsection{Dimension six operator: gluons}

We now compute the Freeze-In density for gluons, keeping in mind the requirement $M_G \lesssim v$, and so $M_G \lesssim T_{R}$. In this section $T_{\rm reh}$ is the reheating temperature.
There should be several regimes of FI. Typical gluons, during FI, will be produced with energy $E_g \sim T_R \gtrsim M_G$. We will focus however only on the case in which the dark sector reaches a gluon plasma.

If the following condition is fulfilled
\bea
\rho^{\rm FI}_{g} \gg \Lambda_{\rm conf}^4 \, ,
\eea 
the dark sector will immediately go back to a plasma of free thermal gluons, then the  
\begin{align}
\dot \rho_g + 4 H \rho_g &= \frac{3T}{512 \pi^6} \int_0^{\infty} d\Omega ds \frac{s}{4} |\mathcal{M}|_{\phi \phi \to gg}^2 K_1 (\sqrt{s}/T) \nonumber \\
&= \frac{3(N^2-1)T}{256 \pi^5 \Lambda^4} \int_0^{\infty} ds s^3 K_1 (\sqrt{s}/T) \\ \nonumber
&=  \frac{4725(N^2-1)T^9}{256 \pi^5 \Lambda^4} \,.
\end{align}
The density normalized to the entropy density, $Y_\psi^{\rm FI}=\frac{n_\psi^{\rm FI}}{s}$ will hence become
\begin{align}
\label{eq:rho_FI}
&\dot \rho_g + 4 H \rho_g = -HT^5\frac{d\left(T^{-4}\rho_g\right)}{dT} = \frac{4725(N^2-1)T^9}{256 \pi^5 \Lambda^4} \nonumber \\
\Leftrightarrow \; &\frac{\rho_g}{T^4}\bigg|_{FI}=\int_0^{\rm T_{reh}} dT  \frac{4725(N^2-1)T^2M_{\rm Pl}}{256 \sqrt{g_\star}1.66\pi^5 \Lambda^4}= 10.2 \frac{(N^2-1)T^3_{\rm reh}M_{\rm Pl}}{ \sqrt{g_\star}\pi^5 \Lambda^4}
\end{align}
 which correspond to a temperature of the dark sector being 
\bea  
\bigg(\frac{T_g}{T_\gamma}\bigg)^4 \approx 150 \frac{T^3_{\rm reh}M_{\rm Pl}}{ \sqrt{g_\star}\pi^7 \Lambda^4} \, . 
\eea

\section{Computation of the instantaneous spectrum after production}
\label{app:spectrum_at_em}

In this appendix, we present the computations of the spectrum of boosted particles immediately after emission. We will first present a few analytical results and then provide the numerical method we followed for the full computation of the spectrum that appears in the main text.

\subsection{Energy distributions}
Let us calculate the energy distribution for the particles in  the plasma-bubble wall collisions. We will now compute the energy spectrum immediately after the DM production from bubble expansion.

\subsubsection{Dimension 5 operator computation}
We will start the discussion with the case of the dimension five operator. The average value of the transverse momenta  of the field $\psi$ will be given by, for an incoming $h$ particle with fixed momentum $p$,
\bea 
\bar k_\perp \equiv  \frac{\int d^3p P_{h \to \psi \psi} f_h(p) k_\perp}{\int d^3p P_{h \to \psi \psi} f_h(p)}, \qquad \bar k^2_\perp \equiv  \frac{\int d^3p P_{h \to \psi \psi} f_h(p) k^2_\perp}{\int d^3p P_{h \to \psi \psi} f_h(p)} \, ,
\eea 
which reduce to 
\bea
\label{eq:average_mom_5}
\bar k_\perp=\frac{\int \frac{d  k_\perp^2 dx }{(k_\perp^2+M_\psi^2)^2} \l[k_\perp^2+{M_\psi^2}(2 x-1)^2\r] k_\perp}{\int \frac{d  k_\perp^2 dx }{(k_\perp^2+M_\psi^2)^2} \l[k_\perp^2+{M_\psi^2}(2 x-1)^2\r] }\qquad \text{with} \quad
\frac{k_\perp^2+M_\psi^2}{2 p_0 x (1-x)}< L_w^{-1}
\nn
\Rightarrow \qquad 
\bar k_\perp=\frac{\pi \sqrt{p_0 L_w^{-1}}}{2\sqrt 2 (\log \frac{p_0  L_w^{-1}}{M_\psi^2}-8/3+\log2)}
,~~
\qquad 
\bar k_\perp^2\simeq \frac{L_w^{-1} p_0}{3 \log \big(\frac{p_0 L_w^{-1}}{M_\psi^2}\big)-5.92
} \,.
\eea

Armed with this expression we can estimate the average energy of the $\psi$ field in the plasma frame $\bar{E_\psi}_{\rm plasma}$, assuming that the incoming particle $h$, which produces $\psi$, has energy $p_0$ $\sim p_z$. We obtain
\bea
\label{eq:energy-plasma-frame}
\bar{E_\psi}_{\rm plasma}= \gamma_w (E_\psi-v_w \sqrt{E_\psi^2 -k_\perp^2-M_\psi^2}) \approx \gamma_w \frac{\bar k_\perp^2}{2E_\psi}\,,
\eea
where $E_\psi$ is the energy of the $\psi$ field in the wall frame and scales like $\sim T\gamma_w$. In this case, we can expand for very large values of $\gamma_w$ and use the expression for the average value of $\bar k_\perp^2$ to find
\bea 
\label{eq:average_E}
\bar E_\psi\simeq \frac{L_w^{-1} \gamma_w}{3\log \frac{\gamma_w T}{M_\psi ^2 L_w} -5.92} \,, 
\eea 
where we took $2E_\psi \approx p_0$, corresponding to $x \approx 1/2$. This analytical computation does not take into account the necessary convolution with the Boltzmann distribution of the incoming $h$.

\subsubsection{Dimension 6 operator computation}
For the dimension six case, the computation proceeds in a similar way.
We can again neglect the mass of the $M_\gamma$ in the computation and the expressions simplify to  
\begin{align}
\bar{k}_{\perp}^2\simeq \frac{p_0 (17-24 \log (2))}{6 L_w (\log (4)-1)}\approx 0.16 \; p_0L_w^{-1}. 
\end{align} 
\begin{align}
\bar{k}_{\perp}\simeq -\frac{\left(16 \sqrt{2}-23\right) \pi  p_0}{6 \sqrt{2} (\log (4)-1) \sqrt{L_w p_0}}\approx 0.36 \sqrt{p_0 L_w^{-1}} 
\end{align}

The average energy in the plasma frame will then become
\begin{align}
\label{eq:photons_Aver_E}
\bar{E_\g}_{\rm plasma}^{\rm analytical} &\approx \gamma_w \frac{\bar k_\perp^2}{2E_\g}  \approx 0.16 \: \gamma_w L_w^{-1}\,, 
\end{align}

where we again took $2E_\gamma \approx p_0$ since $x\approx \frac{1}{2}$.

\subsection{Numerical algorithm for the evaluation of the spectrum}

We have also calculated numerically the distribution function with respect to the energy of the $\psi$ field. This can be done by convoluting the distribution of the initial particle with the corresponding $\delta$ function
\bea
&&\frac{d F}{d E}=\frac{\int d^3 p f_h(p) \int d  k_\perp^2 dx |\mathcal{M}_{h \to XX }|^2 \delta (E-E(p^h_0,p_z,x,k_\perp))}{\hbox{Normalization}}\nn
&&E(E,p_z,x,k_\perp))=p_0^h x \gamma_w-\sqrt{\gamma_w^2-1}\sqrt{(p^h_0x)^2-M^2-k_\perp^2},
\eea
where the integral is performed using the Monte-Carlo method and $E$ ($M$) is the energy (mass) of the emitted particle, being $E_\psi, E_\gamma$ ($M_\psi, M_\gamma$). The overall normalization factor is determined at the end requiring
\bea
\label{eq:norm}
\int \frac{d F}{d E} d E=1 \,. 
\eea
The numerical procedure  becomes more efficient if we introduce a new integration variable:
\bea
Y=\frac{\gamma_w}{T}(p^h_0-v p_z^h) ~\Rightarrow \qquad d^3 p f_h(p) \propto  d Y d p_z e^{-Y} \l( \frac{T Y}{\gamma_w}+ v_w p_h\r).
\eea
Note that we have ignored the overall numerical factor in front since the normalization is anyway determined at the end by Eq.\eqref{eq:norm}. The results of this procedure for the spectrum are shown in Fig.\ref{Fig:spectrum_num} and Fig.\ref{Fig:spectrum_num_vectors}, for the fermion and the vector production respectively, and we see that the distribution is peaked around the average energy value $\bar E$ and drops exponentially fast once the threshold $L_w^{-1}\gamma_w$ is passed. One can see the origin of this threshold from the following (expanding Eq.\ref{eq:energy-plasma-frame}):
\bea
\bar{E}_{\rm plasma}\simeq \frac{E}{2\gamma_w}+\frac{k_\perp^2 \gamma_w}{2E} \approx  \frac{k_\perp^2 \gamma_w}{p^h_0} \, ,
\eea
where we took $p_0 \approx 2E $. On the other hand the maximum value of $k_\perp^2$ is 
\bea
k_\perp^2|_{\rm MAX}\sim p^h_0 L_w^{-1} \, ,
\eea
from non-adiabaticity arguments, thus we obtain
\bea
\bar{E}_{\rm plasma}|_{\rm MAX}\simeq  L_w^{-1} \gamma_w , 
\eea
so we expect the sharp drop of the spectrum once the threshold is passed. Using the spectrum, computed via this method, also allows us to compute the average energy in the plasma frame by simple integration of it. 

\subsubsection{Numerical determination of the average energies}

After having computed the spectrum numerically, we can also extract the average energy via 
\bea
\label{eq:norm}
\bar E = \int \frac{d F}{d E} E d E \,. 
\eea 
and we obtain a good fit of the numerical data with
\bea 
\bar{E_\psi}_{\rm plasma}^{\rm numerical} \simeq (0.5-1)\frac{L_w^{-1} \gamma_w}{3\log \frac{\gamma_w T}{M_\psi^2 L_w} -5.92} \, ,
\eea 
for the dimension five production and 
\bea 
\bar{E_\g}_{\rm plasma}^{\rm numerical} \approx (0.12-0.17) \gamma_w L^{-1}_w \,,
\eea 
for the dimension six production. This agrees well with the analytical computations.

\bibliographystyle{JHEP}
{\footnotesize
\bibliography{biblio}}
\end{document}